\documentclass[12pt,a4paper]{article}
\usepackage{float}
\usepackage{multirow}
\usepackage{inputenc}
\usepackage{hyperref}
\usepackage{graphicx}
\usepackage{wrapfig}
\usepackage{amsmath}
\usepackage{xparse}
\usepackage{amsfonts}
\usepackage{amssymb}
\usepackage{relsize} %big math signs
\usepackage[top=1in, bottom=2cm, left=2cm, right=2cm]{geometry}
\usepackage{authblk}
\usepackage{ulem}
\usepackage{physics}
\usepackage{subcaption}
\usepackage{makecell}
\usepackage{xcolor}
\usepackage{quotes}
\NewDocumentCommand{\tens}{t_}
{%
	\IfBooleanTF{#1}
	{\tensop}
	{\otimes}%
}
\NewDocumentCommand{\tensop}{m}
{%
	\mathbin{\mathop{\otimes}\displaylimits_{#1}}%
}
\usepackage{changepage} %to make tables/figures start from a bit left

\usepackage{afterpage}

\usepackage{placeins}

\usepackage{wrapfig}
\usepackage{cite} % for compact citations
\usepackage[nottoc,notlof,notlot]{tocbibind} % to put the bibliography in the ToC
\usepackage{dblfloatfix}
\newcommand{\orcid}[1]{\href{https://orcid.org/#1}{\includegraphics[width=8pt]{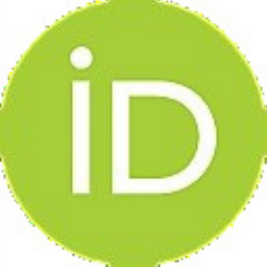}}}

\title{\bf  Noncommutative $p$-wave holographic superconductors}
\author{\bf  \normalsize{Souvik Paul}~\orcid{0009-0002-8919-6275} \thanks{souvik.paul@bose.res.in}}
\author{\bf Sunandan Gangopadhyay~\orcid{0000-0002-0447-9317} \thanks{sunandan.gangopadhyay@gmail.com, sunandan.gangopadhyay@bose.res.in}}

\affil{\large{Department of Astrophysics and High Energy Physics,\\
		S.N.~Bose National Centre for Basic Sciences,\\
		Salt Lake, Kolkata 700106, India}}

\date{}
\begin{document}
	\maketitle
	\begin{abstract}
		\noindent In this work, we have studied the effects of noncommutative geometry on the properties of p-wave holographic superconductors with massive vector condensates in the probe limit. We have applied the St\"{u}rm-Liouville eigenvalue approach to analyse the model. In this model, we have calculated the critical temperature and the value of the condensation operator for two different values of $m^2$. We have also shown how the influence of noncommutative geometry modifies these quantities. Finally, by applying a linearised gauge field perturbation along the boundary direction, we calculated the holographic superconductor's AC conductivity using a self-consistent approach and then carried out a more rigorous analysis. The noncommutative effects are also found to be present in the result of AC conductivity. We have also found that just like the commutative case, here the DC conductivity diverges due to the presence of a first order pole in the frequency regime. 
	\end{abstract}
\section{Introduction}
 In 1911, Kamerlingh Onnes observed that the electrical resistivity of mercury dropped to zero when it was cooled below a temperature of about $4.2$ Kelvin \cite{onnes1911superconductivity,Bardeen:1957mv,tinkham2004introduction}. This remarkable phenomenon indicated that mercury could conduct electricity without any resistance at very low temperatures, leading to the identification of a new state of matter, namely the superconducting state. Later it was found that most of the metals shows this kind of property when the temperature is cooled down below a certain temperature called the critical temperature.\\
 Later in the 1950s, Landau and Ginzburg \cite{ginzburg2009landau} gave a theoretical model for superconductivity. The Landau-Ginzburg theory employs an order parameter, typically denoted as $\psi$, to describe the transition between the normal and superconducting states. The order parameter has a non-zero value when the temperature is below the critical temperature, and it is zero when the temperature is above the critical temperature. This eventually implies the presence of a superconducting and normal phase. In this theory the free energy functional in terms of the order parameter $\psi$ reads
 \begin{equation}
 	F(\psi, T)=\alpha (T-T_{c})|\psi|^{2}+\frac{\beta}{2}|\psi|^{4}+\dots
 \end{equation}
 where $\alpha~,\beta$ are real positive constants. For $T>T_{c}$ the free energy has its minima at $|\psi|=0$ and for $T<T_c$, the free energy has its minima at 
 \begin{equation}
 	|\psi|^{2}=\sqrt{\frac{-2\alpha(T-T_{c})}{\beta}}~.
 \end{equation}
The BCS theory of superconductivity \cite{Bardeen:1957mv} is one of the most successful microscopic theory for describing the properties of low temperature superconductors. The cornerstone of BCS theory is the concept of Cooper pairs, which are formed when two electrons with opposite spin and momentum bind together due to the attractive interactions mediated by phonon. Although when it comes about studying the properties of high temperature superconductors \cite{vandeMarel:2003wn,van2006scaling,Dagotto:1993ajt}, the BCS theory fails as it is known to explain weakly coupled superconductors. \\
Over the past few decades, the AdS/CFT correspondence \cite{Maldacena:1997re,Gubser:1998bc,Witten:1998qj,Aharony:1999ti,Natsuume:2014sfa,Nastase:2007kj} has played a crucial role in exploring strongly coupled quantum field theories by utilising a weakly coupled gravitational theory in AdS space, which includes an additional dimension. In his seminal paper, Juan Maldacena established that the $\mathcal{N}=4$ supersymmetric Yang-Mills theory is dual to a type IIB superstring theory in the space $AdS_{5} \times S^5$. This discovery has gained attention in different branches of physics, starting from condensed matter physics\cite{Sachdev:2010ch,Keski-Vakkuri:2008ffv}, quantum chromodynamics\cite{Csaki:2008dt,Brodsky:2008pf,Erlich:2005qh,Karch:2006pv,Kruczenski:2004me,Aharony:2002up}, quantum gravity \cite{Bak:2006nh,Rovelli:1997na,Engelhardt:2015gla}, cosmology \cite{Nojiri:2000kq,Antonini:2019qkt,PhysRevD.86.043513}, etc.\\
In recent years this strong/weak duality has been used to understand the properties of high $T_c$ superconductors.
Before diving into the holographic model of superconductors, we should know about the rudiments of the phenomena of superconductivity. In the superconducting state, materials show zero resistivity or diverging conductivity. For this, we need a $U(1)$ current. Also superconductors show a phase transition. According to the Ginzburg-Landau theory \cite{ginzburg2009landau}, we need an order parameter to study this phase transition. This order parameter is a charged scalar operator $\langle J\rangle$\footnote{In the literature, this charged scalar operator is sometimes denoted as $\langle O\rangle$. In this paper we represent it by $\langle J\rangle$.}.\\
The holographic model for $s$-wave superconductor consists of an AdS black hole and a complex scalar field minimally coupled to an Abelian gauge field. Below a certain critical temperature, a charged scalar condensate forms outside the black hole which corresponds to the formation of a charged scalar condensate in the dual CFTs. The process of this condensation involves the breaking of a local $U(1)$ symmetry in the vicinity of the black hole's event horizon. In \cite{PhysRevLett.101.031601}, it was proposed that a simple theory of the Abelian Higgs model in anti-de Sitter (AdS) space can lead to spontaneous symmetry breaking, resulting in the emergence of scalar hair near the horizon of a black hole. A more simplified analysis of the $s$-wave holographic superconductor was performed in \cite{Hartnoll:2008kx}, where the probe limit is considered, where the back reactions of the Maxwell gauge field and charged complex scalar field do not affect the spacetime metric.\\
We now mention how the fields in the bulk holographic superconductor model are mapped with the condensed matter superconductivity quantities. In linear response theory \cite{Kubo:1957mj,Callen:1951vq}, one sees the response of an operator to an external source or perturbation. For example, a magnetic field, gauge potential, and spacetime fluctuation acts as an external source and creates responses to magnetization, charge density and energy-momentum tensor, respectively. According to the  Gubser, Klebanov, Polyakov, Witten (GKPW) relation \cite{Gubser:1998bc,Witten:1998qj}, the partition function on the AdS and CFT sides can be equated. Using the equations of motion of the bulk field, the on-shell bulk action can be recast to a surface term on the AdS boundary. This surface term is identified as the generating functional of the boundary CFT. Thus, a bulk field acts as an external source of a boundary operator. For example bulk scalar fields ($\phi$) couple with boundary condensation operator $J$, bulk gauge fields ($A^{\mu}$) with boundary $U(1)$ current $\mathcal{J_{\mu}}$, and gravitational field ($g_{\mu\nu}$) with boundary energy momentum tensor ($T^{\mu\nu}$). In all of these cases, the boundary value of these bulk operators acts as the source for the boundary operators \footnote{For a more detailed understanding of these mappings, the reader may look into these references \cite{Natsuume:2014sfa,Nastase:2007kj,Herzog:2009xv,Musso:2013vtg}. .}The mapping between condensed matter superconductivity quantities and gravity is as follows. As mentioned earlier, superconductivity is a phenomenon associated with a second-order phase transition. Hence, it has an order parameter which is the value that the charged scalar operator takes below a critical temperature, namely, $\left<J\right>$. This corresponds to the macroscopic wave function in the Ginzburg-Landau theory \cite{ginzburg2009landau}. The simplest gravity setup (holographic setup) that can give rise to these ingredients present in superconductivity turns out to be an Einstein-Maxwell-complex scalar system, where the bulk metric of the Einstein-Maxwell black hole geometry corresponds to the energy-momentum tensor $T_{\mu\nu}$ at the boundary, the Maxwell gauge field corresponds to the boundary current, and the complex scalar field in the bulk spacetime corresponds to the condensate $\left<J\right>$ in the boundary which is non-zero below a certain critical temperature. The Hawking temperature of the black hole in the bulk corresponds to the temperature of the boundary conformal field theory \cite{Witten:1998qj,Nastase:2007kj,Natsuume:2014sfa,Hartnoll:2008kx}.
\\
A lot of other studies have been carried out in this direction for Maxwell electrodynamics \cite{Lee:2008xf,Liu:2009dm,Cai:2011ky,Herzog:2009xv,Horowitz:2008bn,Brihaye:2010mr,Siopsis:2010uq,Koutsoumbas:2009pa,Salvio:2012at}. The studies of $s$-wave holographic superconductor was not only restricted to Maxwell electrodynamics, the properties of $s$-wave holographic superconductor was also studied for various non-linear electrodynamics models with usual Einstein gravity and gravity theories with higher curvature corrections also  \cite{Jing:2010cx,Jing:2010zp,Pan:2011vi,Jing:2011vz,Gangopadhyay:2012am,Gangopadhyay:2012np,Ghorai:2017ses,Li:2011xja,Sheykhi:2016meb,Ghotbabadi:2018ahu,Nam:2019sjv,Salahi:2016mbx,Mohammadi:2018hxc,Kruglov:2018jee}. In particular, the Born-Infeld theory of electrodynamics \cite{Born:1934gh,Born:1933qff,born1934quantum,dirac1962extensible}, which is a non-linear model of electrodynamics, was studied in the context of holographic superconductors in \cite{Gangopadhyay:2012am,Gangopadhyay:2012np}. Numerical and analytical studies incorporating the graviton mass have also been carried out in \cite{Zeng:2014uoa,Nam:2020bfq,Li:2019qkt,Parai:2022gcc,Hu:2024lkw}. This kind of gravity theories appear in the IR regime of the Einstein gravity. Although due to the mass of graviton these theories suffer from the instability problem \cite{deRham:2010kj}. It is worth noting that studies on holographic superconductors in the presence of a dynamical gauge field can be found in \cite{Domenech:2010nf,Montull:2011im,Montull:2012fy}.   \\
Another significant gravity model involves a charged vector field in the bulk, serving as the vector order parameter that corresponds to the holographic $p$-wave superconductor. A holographic $p$-wave superconductor model with $SU(2)$ Yang-Mills field in the bulk was found in \cite{Gubser:2008wv,Gubser:2008zu,Roberts:2008ns}. In contrast to an $s$-wave holographic superconductor, the emergence of the condensate in this case spontaneously breaks not only the $U(1)$ symmetry but also the $SO(2)$ rotational symmetry in the $x$-$y$ plane. Later on, rigorous analytical and numerical studies was performed for this model \cite{Zeng:2010fs,Gangopadhyay:2012gx,Roychowdhury:2013aua,Akhavan:2010bf,Pal:2022oui,Chen:2012rlw}.\\
Recently the Maxwell-vector model to construct a $p$-wave holographic superconductor has got a lot of attentions. In this model, a complex vector field is non-minimally coupled to a local $U(1)$ gauge field in the bulk \cite{Cai:2015cya,Cai:2013aca,Cai:2013pda,Cai:2013kaa}. This theory is dual to a strongly coupled system, considering a charged vector operator with a global U(1) symmetry on the boundary. A detailed analytical study of this model in the presence of a nonlinear Born-Infeld parameter can be found in \cite{Chaturvedi:2015hra,Srivastav:2019tkr}. It was observed that the presence of the Born-Infeld parameter is actually suppressing the values of critical temperature and condensation operator, therefore making the condensate harder to form. \\
On the other hand, the noncommutativity of spacetime is a significant area in theoretical physics that has garnered considerable research attention in recent times. In the hope of removing short-distance singularities in quantum field theory, this concept was introduced long ago in 1947 by Snyder \cite{Snyder:1946qz}. Although the concept of noncommutative field theory remained unnoticed for a long time by the scientific community till Seiberg and Witten \cite{Seiberg:1999vs} showed that in the low energy limit, open strings attached to D-branes induce noncommutativity on the D-brane. The rules to move from ordinary quantum field theory (QFT) to NCQFT was also discussed by them. In recent studies, noncommutative (NC) inspired Schwarzschild and Reissner–Nordström (RN) black hole spacetimes were derived in \cite{Nicolini:2005vd,Nicolini:2008aj}. In this work, the Einstein equations of general relativity were solved by incorporating the effects of noncommutativity using a smeared matter source. Notably, an intriguing aspect of this black hole solution is the absence of a physical singularity. Thermodynamic properties of noncommutative higher dimensional AdS black holes with non-Gaussian smeared matter distributions is also studied in \cite{Miao:2015npc,Miao:2016ulg}.\\
The study of a holographic superconductor model incorporating an uncharged NC AdS$_4$ black hole in the presence of Maxwell electrodynamics was first done in \cite{Pramanik:2014mya}. Later the effect of noncommutativity on the properties of superconductors was extended for various nonlinear electrodynamics models in higher dimensions \cite{Ghorai:2016qwc,Maceda:2019woa,Pramanik:2015eka,Parai:2018fbg,Pal:2017mqj}.\\
Although to date, the literature lacks studies examining Maxwell $p$-wave holographic superconductors in the context of noncommutativity. In this paper, we investigate the effect of noncommutativity on the Maxwell $p$-wave holographic superconductor model in the probe limit within Maxwell electrodynamics. It should be noted that the probe limit corresponds to the scenario where the matter fields do not back-react on the spacetime geometry. In this paper, we have analytically derived the critical temperature, the value of the condensation operator, and the conductivity for the holographic $p$-wave superconductor model in presence of noncommutative parameter. We have also graphically shown how the critical temperature and condensation operator vary with the black hole mass and the noncommutative parameter. We also find that all of these results gets modified due to the presence of noncommutativity in spacetime.\\
%P wave models\\
This paper is organised as follows: In section \eqref{sec 2}, we have discussed about the set up for the $p$-wave holographic superconductor model in noncommutative spacetime. In section \eqref{sec 3}, we have discussed about the asymptotic behaviour of the matter and the gauge fields. In section \eqref{sec 4}, we have calculated an expression of the critical temperature using St\"{u}rm-Liouville eigenvalue approach. In section \eqref{sec 5}, we have derived an expression for the condensation operator and the critical exponent. In section \eqref{sec 6}, we have calculated the value of AC conductivity using a self-consistent approach and using a generalised approach for two different cases (for $\rho_{+}=0$ and $\rho_{-}=0$). Finally, we conclude in section \eqref{sec 7}.  
\section{$p$-wave holographic superconductor in noncommutative spacetime}\label{sec 2}
We start our analysis by considering a noncommutative charged Schwarzschild-AdS$_4$ black hole background, whose metric is given by 
\begin{equation}
	ds^2 = -f(r) dt^2 + \frac{dr^2}{f(r)}+ r^2 d\Omega^{2}
\end{equation}
where $f(r)$ is the lapse function given by \cite{Nicolini:2005vd,Nicolini:2008aj}
\begin{equation}
	f(r)=K+\frac{r^2}{L^2}-\frac{4MG}{r \sqrt{\pi}}\gamma(\frac{3}{2},\frac{r^2}{4\theta})+\frac{G Q^2}{\pi r^2} \left[\gamma^{2}(\frac{1}{2},\frac{r^2}{4\theta})-\frac{r}{\sqrt{2\theta}}\gamma(\frac{1}{2},\frac{r^2}{2\theta})+\sqrt{\frac{2}{\theta}}\gamma(\frac{3}{2},\frac{r^2}{4\theta})\right] ~.
\end{equation}  
where 
\begin{equation}
    \gamma(s,x)=\int_{0}^{x}t^{s-1}e^{-t}dt
\end{equation}
is the lower incomplete gamma function\footnote{Note that the lower incomplete gamma function $\gamma(s,x)$ and the upper incomplete gamma function $\Gamma(s,x)=\int_{x}^{\infty}t^{s-1}e^{-t}dt$ is related by $\gamma(s,x)+\Gamma(s,x)=\Gamma(s)$, where $\Gamma(s)=\int_{0}^{\infty}t^{s-1}e^{-t}dt$ is the Euler gamma function.}. $Q$ is the total charge of the black hole and $K$ is the curvature. $K=1,0,-1$ represents spherical, planar and hyperbolic black hole geometry. This kind of metric can be obtained using a smeared mass and charge distributions \cite{Nicolini:2005vd}
\begin{align}
	&\rho_{matter}(r)=\frac{M}{(4\pi \theta)^{3/2}}e^{-\frac{r^2}{4\theta}}\nonumber\\
	&\rho_{charge}(r)=\frac{Q}{(4\pi \theta)^{3/2}}e^{-\frac{r^2}{4\theta}}
\end{align}
where $\theta$ is the noncommutative length scale which appears due to noncommutativity of space, that is, $[x,y]=i \theta$. \\
To study the holographic superconductor, we will mainly focus on planar black hole geometry, that is $K=0$. In the probe limit we will neglect the metric's $Q^2$ dependent back-reacting terms. Hence the lapse function is taken to be $f_{1}(r)=\frac{r^2}{L^2}-\frac{4MG}{r \sqrt{\pi}}\gamma(\frac{3}{2},\frac{r^2}{4\theta})$. 
We will now introduce the model action for the p-wave holographic superconductor consisting of a gauge field $A_{\mu}$ and a complex massive vector field $\rho_{\mu}$. Hence the total action of this model is given as

\begin{equation}
	\mathcal{S}=\frac{1}{16\pi G_{4}}\int d^4 x \sqrt{-g}\left(R-2\Lambda+\mathcal{L}\right)\label{action p wave massive}
\end{equation}
where $\mathcal{L}$ is the matter Lagrangian present in the background of the noncommutative Schwarzschild-AdS$_4$ black hole, that is given by \cite{Cai:2013aca,Cai:2015cya,Wen:2018lph}
\begin{equation}
	\mathcal{L}=-\frac{1}{4}F_{\mu\nu}F^{\mu\nu}-\frac{1}{2}\rho_{\mu\nu}\textsuperscript{\textdagger}\rho^{\mu\nu}-m^{2} \rho\textsuperscript{\textdagger}_{\mu}\rho^{\mu}\nonumber+\iota \gamma \rho_{\mu}\rho_{\nu}\textsuperscript{\textdagger}F^{\mu\nu}~.\label{matter lagrangian}
\end{equation}
In the above equation $F_{\mu\nu}=\partial_{\mu} A_{\nu}-\partial_{\nu} A_{\mu}$, $\rho_{\mu\nu}=D_{\mu}\rho_{\nu}-D_{\nu}\rho_{\mu}$, $D=\partial_{\mu}-\iota A_{\mu}$ and $\gamma$ is the magnetic moment of the vector field $\rho_{\mu}$. For simplicity of calculation, we have neglected the effect of the magnetic field on the superconductor transition. \\
By varying the action with respect to $\rho_{\mu}$ and $A_{\mu}$, we obtain the following equations of motion 
\begin{align}\label{EOM rho}
	\partial_{\mu}(\sqrt{-g}\rho^{\mu\nu})-\sqrt{-g}(iA_{\mu}\rho^{\mu\nu}+m^{2}\rho^{\nu})=0~,
\end{align}
\begin{align}\label{EOM F}
	\partial_{\mu}(\sqrt{-g}F^{\mu\nu})+2\iota \sqrt{-g}\rho_{\mu}\textsuperscript{\textdagger}\rho^{\mu\nu} =0~.
\end{align}
As the black hole metric only depends on $r$, we will consider the following ansatz for the massive vector field $\rho_{\mu}$ and the gauge field $A_{\mu}$ 
\begin{equation}
	\rho_{\mu}=\delta^{x}_{\mu}\rho (r)~;~A_{\mu} = \delta^{t}_{\mu}\phi (r)~.
\end{equation}
Now putting this ansatz in eq.(s)(\eqref{EOM rho},\eqref{EOM F}), we obtain the following equations of motions for the matter and gauge field
\begin{eqnarray}
	\rho^{''}(r)+\frac{f^{'}(r)}{f(r)}\rho^{'}(r)+\left(\frac{\phi^{2}(r)}{f^{2}(r)}-\frac{m^2}{f(r)}\right)\rho(r) =0~,\label{fafa}\\
	\phi^{''}(r)+\frac{2}{r}\phi^{'}(r)-\frac{2\rho^{2}(r)}{r^2 f(r)}\phi(r) =0~.\label{rwe}
\end{eqnarray} 	
In the above equations $f(r)=K-\frac{4MG}{r\sqrt{\pi}}\gamma(\frac{3}{2},\frac{r^2}{4\theta})+\frac{r^2}{L^2}$~. As we are only interested in planar holographic superconductors, we will set $K=0$. This parameter $K$ determines whether the black hole geometry is planar, spherical or hyperbolic in nature. The case $K=0$, which we have considered in this paper, corresponds to the planar horizon. In holographic superconductors, $K=0$ is typically used because it corresponds to a flat boundary field theory. This choice of $K$ not only simplifies many analytical calculations but also supports the characteristics of Cuprates \cite{Bednorz:1986tc,Tsuei:2000zz}, a class of high-temperature superconductors, which show superconductivity in layered crystal structures. We will also assume the AdS radius ($L$) to be one. The AdS radius $L$ sets the length scale in the bulk AdS space, and it is related to the cosmological constant. Without the loss of generality this quantity can be rescaled to one. Since the physical quantities can later be reintroduced using dimensional analysis, setting $L=1$ makes the analytical calculations easier. Hence the lapse function becomes $f(r)=r^{2}-\frac{4MG}{r\sqrt{\pi}}\gamma(\frac{3}{2},\frac{r^2}{4\theta})$.\\
With this lapse function in hand, we can now calculate the ADM mass of the noncommutative black hole. Using the fact that at the black hole horizon $r=r_{+}$(horizon radius), the lapse function $f(r)$ vanishes, that is $f(r_+)=0$, allows us to determine the ADM mass of the black hole, which is given by
\begin{equation}\label{ADM mass}
	\frac{r^{3}_{+}}{\gamma(\frac{3}{2},\frac{r^{2}_{+}}{4\theta})}=\frac{4MG}{\sqrt{\pi}}~.
\end{equation}
Now using eq.\eqref{ADM mass}, we can rewrite the lapse function in the following form
\begin{equation}
	f(r)=r^2 - \frac{r^{3}_{+}}{r}\frac{\gamma(\frac{3}{2},\frac{r^{2}}{4\theta})}{\gamma(\frac{3}{2},\frac{r^{2}_{+}}{4\theta})}~.
\end{equation}
The Hawking temperature of the black hole is given by
\begin{align}
	T=\frac{f^{'}(r_{+})}{4\pi}=\frac{1}{4\pi}\left[3r_{+}-r^{2}_{+}\frac{\gamma^{'}(\frac{3}{2},\frac{r^{2}_{+}}{4\theta})}{\gamma(\frac{3}{2},\frac{r^{2}_{+}}{4\theta})}\right]~.
\end{align}
Using the properties of the lower incomplete gamma function, we can further simplify the above expression of the black hole temperature, which reads
\begin{equation}
	T=\frac{r_{+}}{4\pi}\left[3-\frac{4MG}{\Gamma(3/2)}\frac{e^{-r^{2}_{+}/4\theta}}{(4\theta)^{3/2}}\right]~\label{BH temp}
\end{equation}
where $\Gamma(3/2)=\int_{0}^{\infty}t^{1/2}e^{-t}dt$.\\
From the AdS/CFT duality, the Hawking temperature of the black hole is identified as the temperature of the boundary CFT \cite{Maldacena:1997re,Natsuume:2014sfa,Nastase:2007kj,Herzog:2009xv,Zaanen:2015oix}. This can be understood in a way that the thermal excitations of the finite temperature boundary CFT are analogous to the emission of the Hawking radiation quanta from the event horizon of the black hole.
\section{Asymptotic behaviour of $\texorpdfstring{\rho}{rho}$ and $\texorpdfstring{\phi}{phi}$}\label{sec 3}
In this section, we shall discuss 
At the asymptotic boundary ($r\to \infty$), which implies $f(r)\to r^2$ and $f^{'}(r)\to 2r$\footnote{As the noncommutative parameter ($\theta$) is small enough and we are looking at the asymptotic behaviour of the matter and the gauge fields, we can consider $e^{-r^{2}/4\theta}<<1$ .}. Hence, in this asymptotic limit eq.\eqref{fafa} becomes
\begin{equation}
	r^2 \rho^{''}(r)+2r\rho^{'}(r)-m^2 \rho(r) =0~.
\end{equation}
The solution of the above equation is given by
\begin{equation}\label{rho general1}
	\rho(r)=\frac{\rho_{+}}{r^{\Delta_{+}}}+\frac{\rho_{-}}{r^{\Delta_{-}}}
\end{equation}
where $\Delta_{\pm}=\frac{1}{2}\left(1\pm \sqrt{1+4m^2}\right)$ is the conformal dimension of the boundary field theory, which depends upon the mass of the vector field. $\rho_+$ and $\rho_-$ are constants corresponding to the conformal dimensions $\Delta_+$ and $\Delta_-$. We will see in the subsequent sections that $\rho_+$ and $\rho_-$ get identified as the condensate of the massive bulk vector field. From the expression of $\Delta$ it is evident that for $\Delta$ being positive and real, $m^2$ must satisfy  
\begin{equation}
	m^2 \geq -\frac{1}{4}~.
\end{equation}
The above inequality is called the famous Breitenlohner-Freedman (BF) bound \cite{Breitenlohner:1982bm}. This bound tells us that although the mass of the vector field is negative, it is stable in the AdS spacetime.\\
Similarly, in the asymptotic limit eq.\eqref{rwe} becomes
\begin{equation}\label{asymp phi eq}
	r^2 \phi^{''}(r)+2r\phi^{'}(r)=0~.
\end{equation}
Eq.\eqref{asymp phi eq} has the solution having the form
\begin{equation}\label{phi c1 c2}
	\phi(r) = c_1 -\frac{c_2}{r}
\end{equation}
where $c_1$ and $c_2$ are constants of integration. In order to identify these constants $c_1$ and $c_2$, we need to recall the mapping between bulk fields and the CFT operators. We have already mentioned in the introduction that near the AdS boundary, the bulk gauge fields couple with the boundary current, and the boundary value of the gauge field acts as a source to the boundary current. Now applying the results of linear response theory in AdS/CFT duality, previous literature showed that in the near boundary limit, "slow falloff of bulk field" acts as a source and "the fast falloff of bulk field" acts as response to the source \cite{Nastase:2007kj,Musso:2013vtg,Marolf:2006nd,Klebanov:1999tb}. Therefore, from the near-boundary solution of the gauge field in eq.\eqref{phi c1 c2}, we can see $c_1$ is the term that falls off slowly and $c_2$ falls off faster. Thus, from the AdS/CFT dictionary, we can identify $c_1$ to be the source that is the chemical potential ($\mu$), and $c_2$ is the response to it, that is, the charge density ($\tilde{\rho}$).
% From the gauge/gravity duality, one can identify $c_1$ to be the chemical potential ($\mu$) and $c_2$ to be the density ($\tilde{\rho}$) of the holographic superconductor.
Hence we have
\begin{equation}
	\phi(r) = \mu -\frac{\tilde{\rho}}{r}~. \label{phi mu rho}
\end{equation}
Before proceeding further, we will do a change of variables, that is $z=\frac{r_+}{r}$. It should be noted that after this change of variable, the black hole horizon is located at $z=1$ and the AdS boundary is now at $z=0$. In terms of the $z$ coordinate, eq.(s)(\eqref{fafa},\eqref{rwe}) can be written as 
\begin{eqnarray}
	&\rho^{''}(z)+\left(\frac{f^{'}(z)}{f(z)+\frac{2}{z}}\right)\rho^{'}+\frac{r^{2}_{+}}{z^4}\left(\frac{\phi^{2}(z)}{f^{2}(z)}-\frac{m^2}{f(z)}\right)\rho(z)=0\label{rho z}~,\\
	&\phi^{''}(z)-\frac{2\rho^2}{z^{2}f(z)}\phi(z)=0~.\label{phi z}
\end{eqnarray}
\section{Near the critical temperature behaviour of matter and gauge fields}\label{sec 4}
In this section, we will discuss the behaviour of matter and gauge fields near the critical temperature. At the critical temperature $T=T_c$, the massive vector field ($\rho(z)$) vanishes. Hence from eq.\eqref{phi z}, we can get
\begin{equation}
	\phi^{''}(z)=0~.
\end{equation}
The above equation has a solution of the form $\phi(z)=c_3 +c_4 z$. Now to determine the constants $c_3$ and $c_4$, we will use the fact that the gauge field vanishes at the black hole horizon, that is, $\phi (1)=0$.
Hence, using the above condition and eq.\eqref{phi mu rho}, one can get the following expression of the gauge field
\begin{equation}
	\phi(z)=\lambda r_{+c}(1-z) \label{phi in critical}
\end{equation}
where $\lambda=\frac{\tilde{\rho}}{r^{2}_{+}}$ and $r_{+c}$ is the horizon radius at the critical temperature.\\
We can rewrite the lapse function in the following form
\begin{equation}\label{fz recast}
	f(z)=\frac{r^{2}_{+}}{z^2}g(z)
\end{equation}
where $g(z)=1-z^{3}\Xi$ and $\Xi \approx1+\frac{(2M)^{\frac{1}{3}}}{\Gamma(\frac{3}{2})}\frac{e^{-\frac{(2M)^{\frac{2}{3}}}{4 \theta}}}{\sqrt{4\theta}}+\frac{1}{2}\frac{\sqrt{4\theta}}{(2M)^{\frac{1}{3}}\Gamma(\frac{3}{2})}e^{-\frac{(2M)^{\frac{2}{3}}}{4 \theta}}$.
So eq.\eqref{rho z} becomes 
\begin{equation}
	\rho^{''}(z)+\frac{g^{'}(z)}{g(z)}\rho^{'}(z)+\left(\frac{\phi^2}{g^{2}(z)r^{2}_{+}}-\frac{m^2}{g(z)z^2}\right)\rho(z)=0~. \label{rho in g}
\end{equation}
Note that $f(z)$ has been written in the form given in eq.\eqref{fz recast} to recast eq.\eqref{rho z} in the simplified form given in eq.\eqref{rho in g}.\\
Now one can take the profile of the massive vector field as 
\begin{equation}
	\rho(z)=\frac{\left<J\right>}{r^{\Delta}_{+}}z^{\Delta}F(z)\label{ansatz rho}~.
\end{equation}
The reason behind choosing the above form of the massive vector field lies in the fact that in the absence of $\rho_{-}$, $\rho_{+}$ behaves as the response, that is the charged scalar condensate ($\left<J\right>$) of the boundary theory for the source $\rho_{-}$. Also, we know from eq.\eqref{rho general1}, the asymptotic behaviour of the charged massive vector field. Therefore, slightly away from the critical temperature, one should expect slight modifications for the profile of $\rho(z)$, which is done by multiplication of a function $F(z)$.
Following the asymptotic behaviour of the matter field $\rho(z)$, $F(z)$ must have the property $F(0)=1$ and $F^{'}(0)=0$. We would also like to mention that the conformal dimension $\Delta$ can be either $\Delta_{+}$ or it can be $\Delta_{-}$. Now substituting the above ansatz from eq.\eqref{ansatz rho} in eq.\eqref{rho in g} and using the expression of $\phi (z)$ from eq.\eqref{phi in critical}, we get the following equation for $F(z)$
\begin{align}
	&z^{2\Delta}(1-z^{3}\Xi)F^{''}(z)+\left[2\Delta z^{2\Delta -1}(1-z^{3}\Xi)-3\Xi z^{2\Delta +2}\right]F^{'}(z)\nonumber\\&+z^{2\Delta}(1-z^{3}\Xi)\left[\frac{\Delta (\Delta -1)}{z^2}-\frac{3\Xi\Delta z}{(1-z^{3}\Xi)}-\frac{m^2}{z^{2}(1-z^{3}\Xi)}\right]F(z)\nonumber\\&+\frac{\lambda^2}{(1-z^{3}\Xi)}z^{2\Delta}(1-z)^{2}F(z)=0~.
\end{align}
By a little algebra, the above expression can be transformed into the well-known St\"{u}rm-Liouville form, which reads
\begin{align}
	&\frac{d}{dz}\left[z^{2\Delta}(1-z^{3}\Xi)F^{'}(z)\right]+z^{2\Delta}(1-z^{3}\Xi)\left[\frac{\Delta (\Delta -1)}{z^2}-\frac{3\Xi\Delta z}{(1-z^{3}\Xi)}-\frac{m^2}{z^{2}(1-z^{3}\Xi)}\right]F(z)\nonumber\\&+\frac{\lambda^2}{(1-z^{3}\Xi)}z^{2\Delta }(1-z)^{2}F(z)=0~.\label{F in SL form}
\end{align}
It is well known that the St\"{u}rm-Liouville form of a differential equation looks as follows
\begin{equation}\label{SL form}
	\frac{d}{dz}\left(p(z)F^{'}(z)\right)+q(z)F(z)+\lambda^{2}r(z)F(z)=0~.
\end{equation}
Now comparing eq.\eqref{F in SL form} and eq.\eqref{SL form}, we can identify
\begin{eqnarray}
	&p(z)=z^{2\Delta }(1-z^{3}\Xi)\\
	&q(z)=z^{2\Delta }(1-z^{3}\Xi)\left[\frac{\Delta (\Delta -1)}{z^2}-\frac{3\Xi\Delta z}{(1-z^{3}\Xi)}-\frac{m^2}{z^{2}(1-z^{3}\Xi)}\right]\\
	&r(z)=\frac{z^{2\Delta }}{(1-z^{3}\Xi)}(1-z)^2~.
\end{eqnarray}
With the above identifications of $p(z)$, $q(z)$ and $r(z)$, we can find the eigenvalue $\lambda^2$ in eq.\eqref{F in SL form} from the following relation
\begin{equation}\label{lamda square general}
	\lambda^2 = \frac{\int_{0}^{1}dz\left(p(z)F^{'2}+q(z)F^2\right)}{\int_{0}^{1}dz~r(z)F^2}\equiv \frac{I_1}{I_2}~.
\end{equation}
In order to evaluate the value of $\lambda^2$, we will choose a trial function $F_{\alpha}(z)=1-\alpha z^2$. 
With this trial function, the integrals $I_1$ and $I_2$ in eq.\eqref{lamda square general}, are given by
\begin{align}
    I_1&=\int_{0}^{1}dz\left(p(z)F^{'2}+q(z)F^2\right)=\int_{0}^{1}dz\left(p(z)(-2\alpha z)^2+q(z)(1-\alpha z^2)^2\right)\label{I1}\\
    &I_2=\int_{0}^{1}dz~r(z)F^2=\int_{0}^{1}dz~r(z)(1-\alpha z^2)^2\label{I2}~.
\end{align}
The reason behind choosing the specific form of the function $F_{\alpha}(z)$ lies in the fact that $F_{\alpha}(0)=1$ and $F^{\prime}_{\alpha}(0)=0$, and this matches the boundary (black hole horizon and AdS boundary) behavior of the massive vector field $\rho(z)$. Also, it maintains the regularity of $F_{\alpha}(z)$ at the horizon ($z=1$). Various numerical studies also predict that the trial function usually decreases slowly from $1$ at the AdS boundary to $0$ at the horizon \cite{Siopsis:2010uq,Gregory:2009fj}. The concave nature of  $F_{\alpha}(z)=1-\alpha z^2$ in the range $z\in [1,0]$ supports the validity of this trial function. The eigenvalue $\lambda^2$ can be calculated by minimising the above equation with respect to $\alpha$. The value of $\lambda_{\alpha_{min}}$ is used to calculate the critical temperature of the $p$-wave holographic superconductor in noncommutative black hole spacetime \cite{Pramanik:2014mya,Srivastav:2019tkr}. We would like to mention that the above steps are performed using the software MATHEMATICA $13.3$\cite{Mathematica}. The procedure to evaluate the numerical value of $\lambda_{\alpha_{min}}$ is given in an Appendix.\\
% \begin{enumerate}
%     \item Using the ansatz for $F_{\alpha}(z)$ in eq.\eqref{lamda square general}, we will take the first order derivative of $\lambda^2$ with respect to $\alpha$.
%     \item Now the expression for $\frac{d \lambda^2}{d \alpha}$ will be set to zero to find the roots of $\alpha$. These roots are calculated using numerical root finding.
%     \item Among several roots of alpha, we will now need to find the root that minimises the value of $\lambda^2$. This can be done by checking the sign of $\frac{d^2\lambda^2}{d\alpha^2}$. The root of $\alpha$ for which $\frac{d^2\lambda^2}{d\alpha^2}>0$ gives the minimum value corresponding to $\lambda^2$.
%     \item With the required root of $\alpha$ in hand, we can calculate the corresponding minimum value of $\lambda^2$ using eq.\eqref{lamda square general}. 
% \end{enumerate}
\\
\noindent We will now proceed to establish a relationship between the critical temperature and charged density of the holographic superconductor. This can be done by using the value of $\lambda$ in eq.\eqref{BH temp}, this yields
\begin{align}\label{critical temp}
	T_{c}&=\frac{r_{+c}}{4\pi}\left[3-\frac{4MG}{\Gamma(3/2)}\frac{e^{-r^{2}_{+}/4\theta}}{(4\theta)^{3/2}}\right]\nonumber\\&=\frac{1}{4\pi}\left[3-\frac{4MG}{\Gamma(3/2)}\frac{e^{-\frac{(2MG)^{\frac{2}{3}}}{4\theta}}}{(4\theta)^{3/2}}\right]\left(\frac{\tilde{\rho}}{\lambda}\right)^{\frac{1}{2}}=\eta \tilde{\rho}^{\frac{1}{2}}~
\end{align}
where $\eta=\frac{1}{4\pi \sqrt{\lambda}}\left[3-\frac{4MG}{\Gamma(3/2)}\frac{e^{-\frac{(2MG)^{\frac{2}{3}}}{4\theta}}}{(4\theta)^{3/2}}\right]$. In order to derive the above equation, we have used the relation $r_{+c}^2 =\frac{\Tilde{\rho}}{\lambda}$.\\
We would like to mention that in the above equation, we have used $r^{2}_{+} \approx (2MG)^{\frac{2}{3}}$ in the term $e^{-r^{2}_{+}/4\theta}$. This expression of the critical temperature is for a four dimensional $p$-wave holographic superconductor in the non-commutative black hole spacetime. It is clear that the effect of noncommutativity appears not only in the coefficient of $\frac{1}{4\pi}$ but also enters the expression of critical temperature through $\lambda^2$.\\
 To move ahead we will choose some specific values of the conformal dimension to calculate the critical temperature of the holographic superconductor.
 \subsection*{Case I ($m^2 =0$ and $\Delta_{+} =1$)} 
 The value of the functions $p(z)$, $q(z)$ and $r(z)$ for $m^2 =0$ and $\Delta_{+}=1$ is given by
 \begin{align}\label{pqr case1}
 	&p(z)=z^{2}(1-z^{3}\Xi)\nonumber\\
 	&q(z)=-3\Xi z^{3}\nonumber\\
 	&r(z)=\frac{z^2}{(1-z^{3}\Xi)}(1-z)^{2}~.
 \end{align}
 \begin{center}
 	\captionof{table}{Value of $\frac{T_{c}}{\Tilde{\rho}^{1/2}}$ for $m^2 =0$ and $\Delta_{+}=1$ using the St\"{u}rm-Liouville approach}
 	\begin{tabular}{ |c|c|c|c|c|c|c| } 
 		\hline
 		$\theta$ & $M=10/G_{4}$ & $M=20/G_{4}$ & $M=30/G_{4}$ & $M=40/G_{4}$ & $M=50/G_{4}$\\
 		\hline 
 		$\theta=0.3$ & 0.1208 & 0.1237 & 0.12392 & 0.12393 & 0.12396 \\ 
 		$\theta=0.5$ & 0.1072 & 0.1201 & 0.1230 & 0.12369 & 0.1238\\ 
 		$\theta=0.7$ & 0.0951 & 0.1115 & 0.1188 & 0.12182 & 0.12308\\ 
 		$\theta=0.9$ & 0.0891 & 0.1025 & 0.1122 & 0.11759 & 0.1204\\ 
 		\hline
 	\end{tabular}
 	\label{tab 1}
 \end{center}
We will now provide a side by side comparison between the values of $\frac{T_{c}}{\Tilde{\rho}^{1/2}}$ (for $M=50/G_4$) in Table \eqref{tab 1} with the $\frac{T_{c}}{\Tilde{\rho}^{1/3}}$ values of noncommutative $s$-wave model results (for $M=50/G_5$) in \cite{Ghorai:2016qwc} for $m^2=-3$, $\Delta_+ =3$ and $d=5$ in Table \eqref{table 2 compare}. For both the column in Table \eqref{table 2 compare}, we can see that the values of $\frac{T_{c}}{\Tilde{\rho}^{1/2}}$ and $\frac{T_{c}}{\Tilde{\rho}^{1/3}}$ decreases for increasing values of the noncommutative parameter $\theta$ implying that the condensation is harder to form. In \cite{Pramanik:2014mya}, the same noncommutative $s$-wave model is studied for (3+1)-bulk spacetime dimensions and conformal dimension of charged scalar field $\Delta_+ =1$. It was shown that for black hole mass $M=\frac{40\sqrt{\theta}}{G_4}\approx \frac{28.28}{G_4}$ and noncommutative parameter $\theta=0.5$, the ratio of the critical temperature and square root of charge density ($\frac{T_{c}}{\Tilde{\rho}^{1/2}}$) equals $0.225$. We can compare this value with our obtained result in Table \eqref{tab 1} for the black hole mass $M=30/G_4$, which is given by $0.123$. This shows thatthe  values are quite different for the noncommutative $s$-wave and $p$-wave massive vector field models. 
\begin{table}[h!]
\centering
\begin{tabular}{|c|c||c|}
    \hline
    \multicolumn{1}{|c|}{} & \makecell{\small Our results for $m^2=0$, \\ $\Delta_+ =1$, and $d=4$}
                            & \makecell{\small $s$-wave model results for $m^2=-3$, \\$\Delta_+ =3$ and $d=5$ in \cite{Ghorai:2016qwc}} \\
    \hline
    $\theta$ & $M = 50/G_4$ & $M = 50/G_5$ \\
    \hline
    0.3 & 0.12396 & 0.1946 \\
    0.5 & 0.1238 & 0.1798 \\
    0.7 & 0.12308 & 0.1615 \\
    0.9 & 0.1204 & 0.1505 \\
    \hline
\end{tabular}
\caption{Comparison the values $\frac{T_{c}}{\Tilde{\rho}^{1/2}}$ of our noncommutative $p$-wave massive vector field model for $m^2=0$, $\Delta_+ =1$, and $d=4$ with the $\frac{T_{c}}{\Tilde{\rho}^{1/3}}$ values of noncommutative $s$-wave model results in \cite{Ghorai:2016qwc} for $m^2=-3$, $\Delta_+ =3$ and $d=5$.}
\label{table 2 compare}
\end{table}

 \subsection*{Case II ($m^2 =-3/16$ and $\Delta_{+} =3/4$)} 
  The value of the functions $p(z)$, $q(z)$ and $r(z)$ for $m^2 =-3/16$ and $\Delta_{+}=3/4$ is given by
 \begin{align}
 	&p(z)=z^{3/2}(1-z^{3}\Xi)\nonumber\\
 	&q(z)=\frac{z^{-1/2}}{16}(9z^{3}\Xi -36\Xi z -6)\nonumber\\
 	&r(z)=\frac{z^{3/2}}{(1-z^{3}\Xi)}(1-z)^{2}~.
 \end{align}
  \begin{center}
  	\captionof{table}{Value of $\frac{T_{c}}{\Tilde{\rho}^{1/2}}$ for $m^2 =-3/16$ and $\Delta_{+}=3/4$ using St\"{u}rm-Lioville approach }
 	\begin{tabular}{ |c|c|c|c|c|c|c| } 
 		\hline
 		$\theta$ & $M=10/G_{4}$ & $M=20/G_{4}$ & $M=30/G_{4}$ & $M=40/G_{4}$ & $M=50/G_{4}$\\
 		\hline 
 		$\theta=0.3$ & 0.09722 & 0.09958 & 0.09971 & 0.099718 & 0.09972 \\ 
 		$\theta=0.5$ & 0.08606 & 0.09661 & 0.09896 & 0.09952 & 0.09966\\ 
 		$\theta=0.7$ & 0.07576 & 0.08969 & 0.09562 & 0.09801 & 0.09898\\ 
 		$\theta=0.9$ & 0.06978 & 0.08216 & 0.09019 & 0.09458 & 0.09691\\ 
 		\hline
 	\end{tabular}
 	\label{tab 2}
 \end{center}
\begin{figure}[!h]
	\centering
	\begin{subfigure}[t]{0.5\textwidth}
		\centering
		\includegraphics[width=9cm]{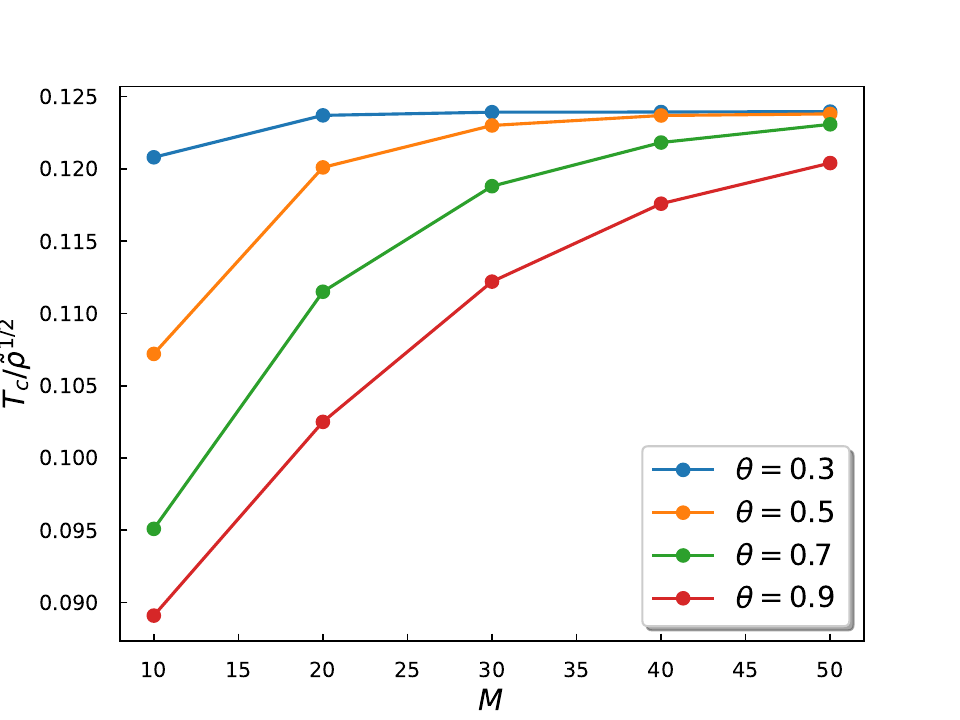}
		\caption{$m^2=0$ and $\Delta_{+}=1$}
		\label{Tc del plus 1}
	\end{subfigure}%
	~
	\begin{subfigure}[t]{0.5\textwidth}
		\centering
		\includegraphics[width=9cm]{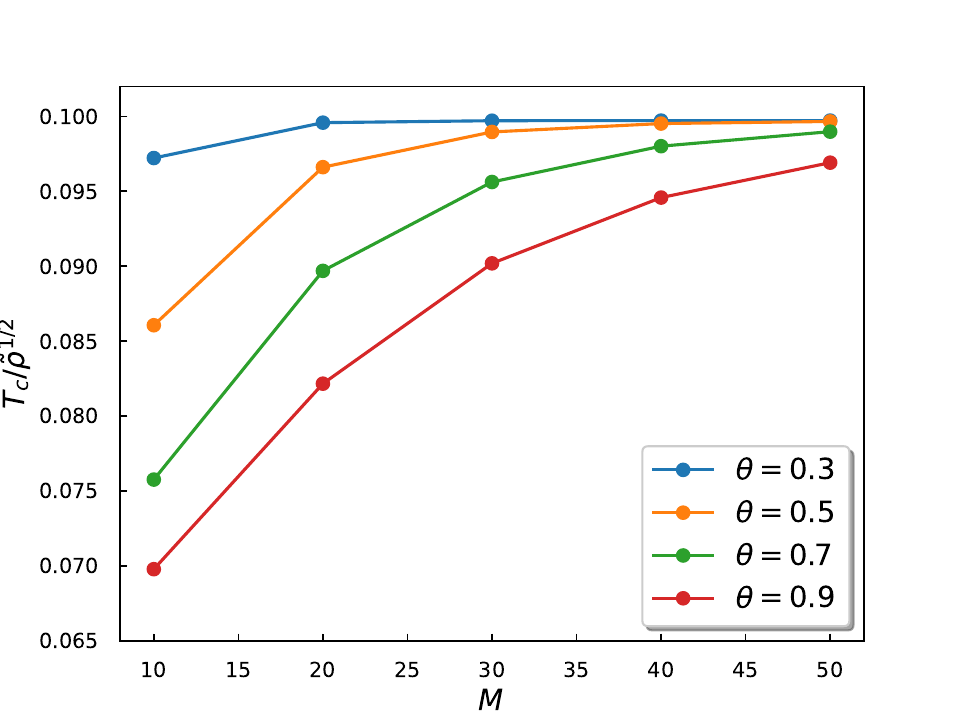}
		\caption{$m^2=-3/16$ and $\Delta_{+}=\frac{3}{4}$}
		\label{Tc del plus 3/4}
	\end{subfigure}
	\caption{Variation of $\frac{T_{c}}{\Tilde{\rho}^{1/2}}$ with black hole mass for different values of noncommutative parameter. In the left panel of the above Figure we have shown the variation of $\frac{T_{c}}{\Tilde{\rho}^{1/2}}$ with black hole mass for $m^2 =0$ and $\Delta_{+}=1$, we have done the plots for different values of the noncommutative parameter. In the right panel, we have done the same plots for $m^2 =-3/16$ and $\Delta_{+}=3/4$}
\end{figure}
 \subsection*{Case III ($m^2 =0$ and $\Delta_{-} =0$)}
  The value of the functions $p(z)$, $q(z)$ and $r(z)$ for $m^2 =0$ and $\Delta_{-}=0$ is given by
 \begin{align}
 	&p(z)=(1-z^{3}\Xi)\nonumber\\
 	&q(z)=0\nonumber\\
 	&r(z)=\frac{(1-z)^2}{(1-z^{3}\Xi)}~.
 \end{align}
 \begin{center}
 	\captionof{table}{Value of $\frac{T_{c}}{\Tilde{\rho}^{1/2}}$ for $m^2 =0$ and $\Delta_{-}=0$ using St\"{u}rm-Lioville approach}
 	\begin{tabular}{ |c|c|c|c|c|c|c| } 
 		\hline
 		$\theta$ & $M=10/G_{4}$ & $M=20/G_{4}$ & $M=30/G_{4}$ & $M=40/G_{4}$ & $M=50/G_{4}$\\
 		\hline 
 		$\theta=0.3$ & 0.08160 & 0.08326 & 0.08336 & 0.08337 & 0.08337 \\ 
 		$\theta=0.5$ & 0.07435 & 0.08119 & 0.08281 & 0.08322 & 0.08333\\ 
 		$\theta=0.7$ & 0.06790 & 0.07666 & 0.08052 & 0.08214 & 0.08283\\ 
 		$\theta=0.9$ & 0.06452 & 0.07189 & 0.07699 & 0.07983 & 0.08139\\ 
 		\hline
 	\end{tabular}
 	\label{tab 3}
 \end{center} 
 \subsection*{Case IV ($m^2 =-3/16$ and $\Delta_{-} =1/4$)} 
  The value of the functions $p(z)$, $q(z)$ and $r(z)$ for $m^2 =-3/16$ and $\Delta_{-}=1/4$ is given by
  \begin{align}
 	&p(z)=z^{1/2}(1-z^{3}\Xi)\nonumber\\
 	&q(z)=-\frac{9}{16}\Xi z^{3/2}\nonumber\\
 	&r(z)=\frac{z^{1/2}}{(1-z^{3}\Xi)}(1-z)^{2}~.
 \end{align}
 \begin{center}
 	\captionof{table}{Value of $\frac{T_{c}}{\Tilde{\rho}^{1/2}}$ for $m^2 =-3/16$ and $\Delta_{-}=1/4$ using St\"{u}rm-Liouville approach}
 	\begin{tabular}{ |c|c|c|c|c|c|c| } 
 		\hline
 		$\theta$ & $M=10/G_{4}$ & $M=20/G_{4}$ & $M=30/G_{4}$ & $M=40/G_{4}$ & $M=50/G_{4}$\\
 		\hline 
 		$\theta=0.3$ & 0.22032 & 0.22584 & 0.22612 & 0.22615 & 0.22615 \\ 
 		$\theta=0.5$ & 0.19392 & 0.21889 & 0.22439 & 0.22569 & 0.22602\\ 
 		$\theta=0.7$ & 0.16958 & 0.20251 & 0.21656 & 0.22216 & 0.22444\\ 
 		$\theta=0.9$ & 0.15562 & 0.18468 & 0.20372 & 0.21409 & 0.21959\\ 
 		\hline
 	\end{tabular}
 	\label{tab 4}
 \end{center}
\begin{figure}[!h]
	\centering
	\begin{subfigure}[t]{0.5\textwidth}
		\centering
		\includegraphics[width=9cm]{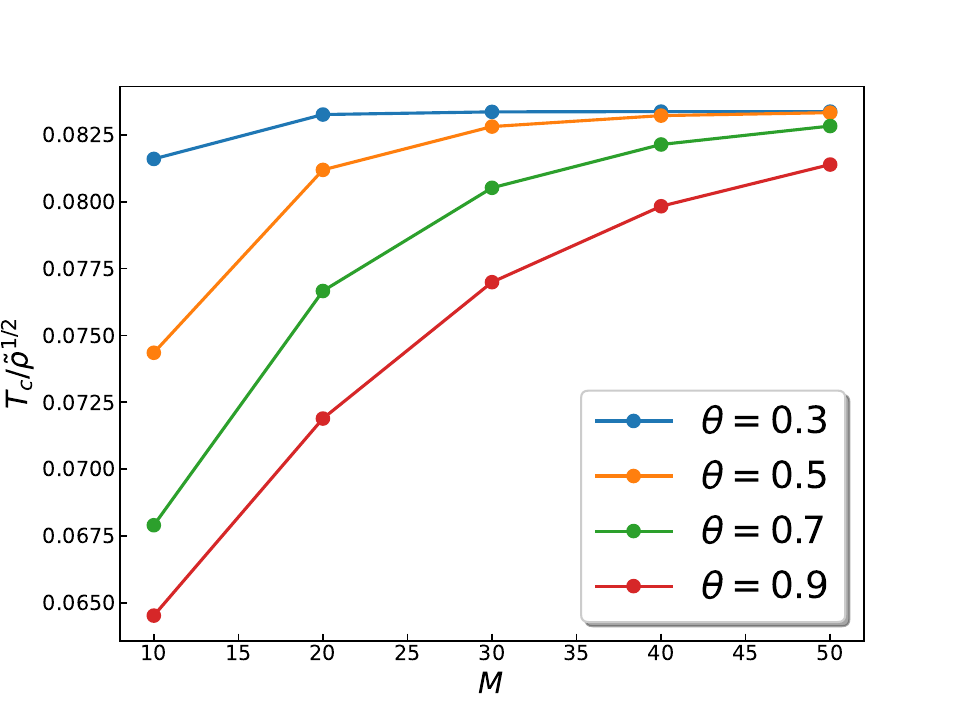}
		\caption{$m^2=0$ and $\Delta_{-}=0$}
		\label{Tc del minus 0}
	\end{subfigure}%
	~
	\begin{subfigure}[t]{0.5\textwidth}
		\centering
		\includegraphics[width=9cm]{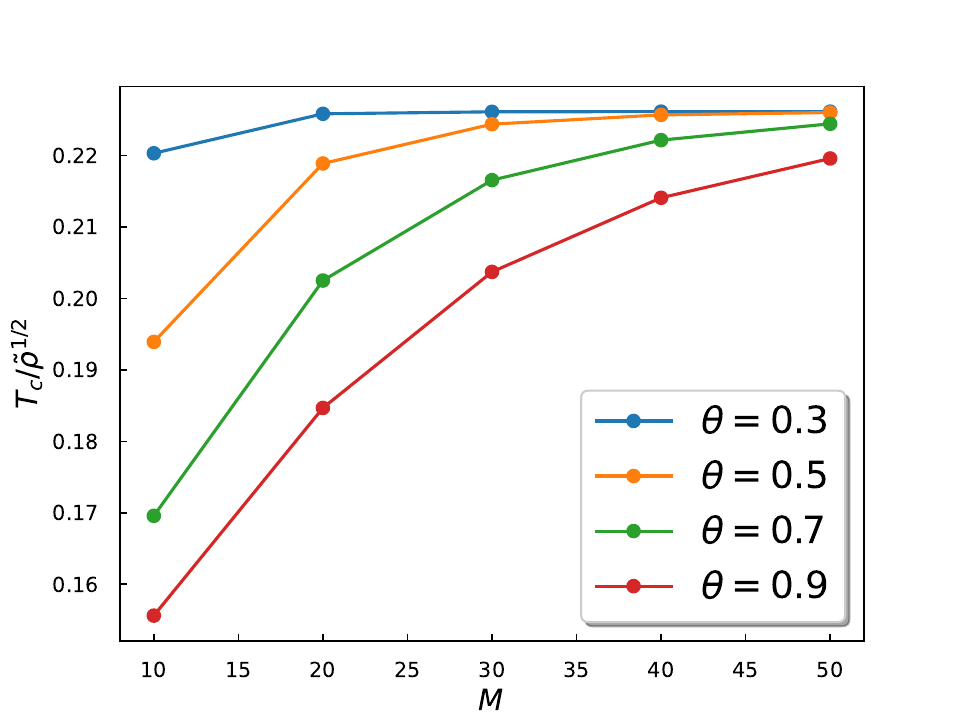}
		\caption{$m^2=-3/16$ and $\Delta_{-}=\frac{1}{4}$}
		\label{Tc del minus 1/4}
	\end{subfigure}
	\caption{Variation of $\frac{T_{c}}{\Tilde{\rho}^{1/2}}$ with black hole mass for different values of noncommutative parameter. In the left panel of the above Figure we have shown the variation of $\frac{T_{c}}{\Tilde{\rho}^{1/2}}$ with black hole mass for $m^2 =0$ and $\Delta_{-}=0$, we have done the plots for different values of the noncommutative parameter. In the right panel, we have done the same plots for $m^2 =-3/16$ and $\Delta_{-}=1/4$}
\end{figure}
In table(\eqref{tab 1},\eqref{tab 2},\eqref{tab 3} and \eqref{tab 4}), we have shown how the values of $\frac{T_{c}}{\Tilde{\rho}^{1/2}}$ changes for different values of noncommutative parameter ($\theta$) and black hole mass (M). For different values of the conformal dimensions (that is $\Delta_{+}=1,3/4$ and $\Delta_{-}=0,1/4$) we have used the corresponding expressions of $p(z)$, $q(z)$ and $r(z)$ to calculate $\frac{T_{c}}{\Tilde{\rho}^{1/2}}$ using the St\"{u}rm-Liouville eigenvalue method. We have also graphically illustrated how the quantity $\frac{T_{c}}{\Tilde{\rho}^{1/2}}$ varies with the black hole mass ($M$) for various values of the noncommutative parameter.\\
In Fig.\eqref{Tc del plus 1}, we have graphically shown the variation of $\frac{T_{c}}{\Tilde{\rho}^{1/2}}$ with respect to the black hole mass ($M$) for $\Delta_{+}=1$. Similarly in Fig.\eqref{Tc del plus 3/4} the graphical variation between $\frac{T_{c}}{\Tilde{\rho}^{1/2}}$ and $M$ is shown for $\Delta_{+}=3/4$. Both of these plots are done for four different values of the noncommutative parameter ($\theta$). In both the figures, the plots corresponding to the colours blue, orange, green and red refer to $\theta=0.3,~0.5,~0.7$ and $0.9$ respectively. It is also clear from these plots that the condensation becomes harder to form for higher values of the noncommutative parameter. We would also like to mention that for higher values of the black hole mass, all the curves approach towards a fixed value of $\frac{T_{c}}{\Tilde{\rho}^{1/2}}$. This means that when the black hole mass is much higher compared to the noncommutative parameter, the effect of noncommutativity is negligible, and it does not affect the properties of the holographic superconductor.\\
In Fig.\eqref{Tc del minus 0}, we have graphically shown the variation of $\frac{T_{c}}{\Tilde{\rho}^{1/2}}$ with respect to the black hole mass ($M$) for $\Delta_{-}=0$. Similarly in Fig.\eqref{Tc del minus 1/4} the graphical variation between $\frac{T_{c}}{\Tilde{\rho}^{1/2}}$ and $M$ is shown for $\Delta_{-}=1/4$. Both of these plots are done for four different values of the noncommutative parameter ($\theta$). In both the Figures the plots corresponding to the colours blue, orange, green and red refers to $\theta=0.3,~0.5,~0.7$ and $0.9$ respectively. Therefore, in this scenario, when we assume $\rho_{+}=0$, it is evident from these plots that the formation of condensation becomes more difficult as the noncommutative parameter increases. Here also, we would like to emphasise that for larger black hole masses, all curves tend to converge to a constant value of $\frac{T_{c}}{\Tilde{\rho}^{1/2}}$. This indicates that when the black hole mass significantly exceeds the noncommutative parameter, the impact of noncommutativity becomes minimal and does not influence the characteristics of the holographic superconductor.\\
From the AdS/CFT point of view, the region close to the horizon is refered to as the infrared (IR) region. It is well known that noncommutativity introduces a physical length scale, below which matter fields cannot be localised. The presence of noncommutativity smears the black hole mass and charge, thereby affecting the near horizon black hole properties, which is the IR domain of the theory. This indicates $\theta$ effectively introduces an IR cutoff in the theory. This IR cutoff smears the gauge and matter fields close to the horizon. As a result, it also affects the chemical potential and charge density of the holographic superconductor, which suggests that the critical temperature will be affected by noncommutativity. One can clearly see from the graphs in Fig.(s)(\eqref{Tc del plus 1},\eqref{Tc del plus 3/4},\eqref{Tc del minus 0},\eqref{Tc del minus 1/4}) that the increasing value of the noncommutative parameter ($\theta$) supresses the value of the ratio of the critical temperature and the square root of the charge density, that is, $\frac{T_{c}}{\Tilde{\rho}^{1/2}}$.
\section{Condensation operator and critical exponent}\label{sec 5}
In this section, we will derive an expression of the condensation operator for a superconductor at the AdS boundary, which is described by the bulk action in eq.\eqref{action p wave massive} in the gravity side. To evaluate this quantity, we must study the properties of the matter and gauge fields slightly away from the critical temperature. At the critical temperature, the behaviour of the gauge field ($\phi (z)$) is governed by the eq.\eqref{phi in critical}. Hence, slightly away from the critical temperature, one must expect that the expression of $\phi (z)$ would slightly differ from eq.\eqref{phi in critical}. Hence we will add a small fluctuation $\chi$ with eq.\eqref{phi in critical} to represent $\phi (z)$ away from the critical temperature, which reads
\begin{equation}
	\phi (z)=\lambda r_{+}(1-z)+\frac{\left<J\right>^2}{r^{2\Delta}_{+}}\chi (z)\label{phi away critical}
\end{equation}   
where $\chi (1)=0$ and $\chi^{'}(1)=0$.\\
To proceed further we will substitute the expression of $\rho (z)$ from eq.\eqref{ansatz rho} in eq.\eqref{phi z}, which leads to the following differential equation for $\phi (z)$
\begin{equation}
	\phi ^{''}(z)=\frac{\left<J\right>^2}{r^{2}_{+}}\mathcal{B}(z)\phi (z) \label{phi J}
\end{equation}
where $\mathcal{B}(z)=\frac{2F^{2}(z)}{f(z)}\frac{z^{2\Delta_{+}-2}}{r^{2\Delta_{+}-2}_{+}}$~.\\
Now differentiating eq.\eqref{phi away critical} twice with respect to $z$ gives us
\begin{equation}
	\phi ^{''}(z)= \frac{\left<J\right>^2}{r^{2}_{+}}\frac{\chi ^{''}(z)}{r^{2\Delta -2}_{+}}~.\label{phi B}
\end{equation}
In eq.(s)(\eqref{phi J},\eqref{phi B}), comparing the coefficients of $\frac{\left<J\right>^2}{r^{2}_{+}}$ gives us the following relation
\begin{equation}
	\chi ^{''}(z)=\lambda r_{+}\frac{2F^{2}(z)}{f(z)}z^{2\Delta -2}(1-z)~.
\end{equation}
We will see in a moment that the above equation for the small fluctuation $\chi$ is very useful in determining the condensation operator. Integrating the above equation from $z=0$ to $z=1$ leads to
\begin{align}
	&\chi ^{'}(z)\bigl\lvert _{0}^{1}=\lambda r_{+}\int_{0}^{1}\frac{2F^{2}(z)}{f(z)}z^{2\Delta -2}(1-z)~dz\nonumber\\
	&\implies \chi ^{'}(0)=-\lambda r_{+}\mathcal{A}_{\Delta}\label{chi prime zero}
\end{align}
where $\mathcal{A}_{\Delta}= \int_{0}^{1}\frac{2F^{2}(z)}{f(z)}z^{2\Delta -2}(1-z)~dz$. While deriving the above expression, we have used the fact that $\chi ^{'}(1)=0$~. Expanding the small fluctuation $\chi (z)$ near the AdS boundary ($z=0$), we obtain
\begin{equation}
	\chi (z)=\chi (0)+z\chi ^{'}(0)+\dots~. \label{chi expand}
\end{equation}
\begin{figure}[!h]
	\centering
	\begin{subfigure}[t]{0.5\textwidth}
		\centering
		\includegraphics[width=8.8cm]{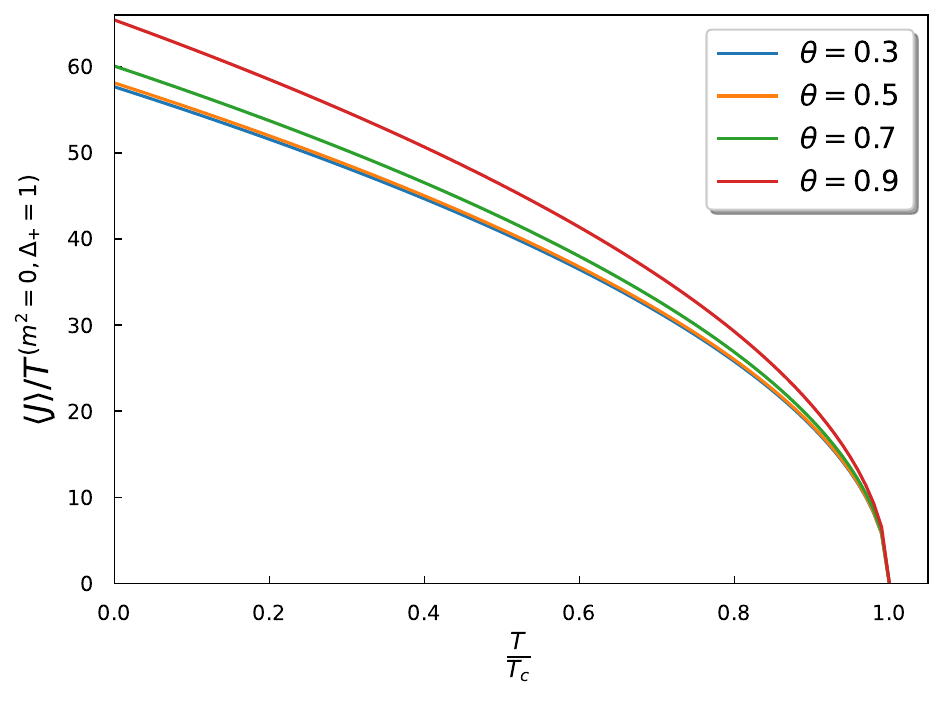}
		\caption{$m^2=0$ and $\Delta_{+}=1$}
		\label{con del plus 1}
	\end{subfigure}%
	~
	\begin{subfigure}[t]{0.5\textwidth}
		\centering
		\includegraphics[width=8.8cm]{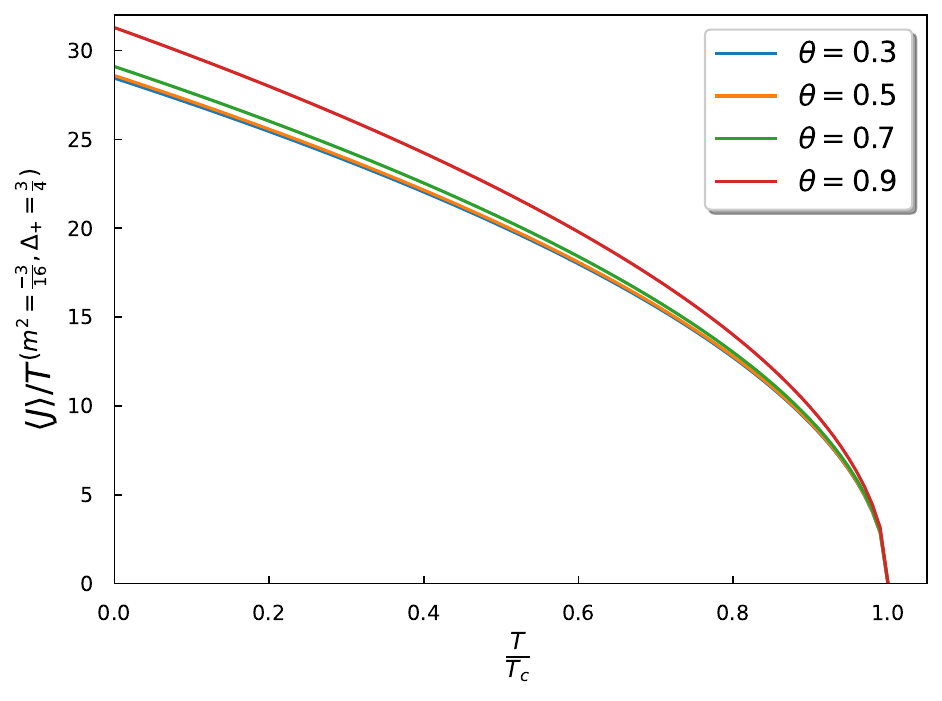}
		\caption{$m^2=-3/16$ and $\Delta_{+}=\frac{3}{4}$}
		\label{con del plus 3/4}
	\end{subfigure}
	\caption{Variation of the dimensionless condensate $\frac{\left<J\right>}{T^{\Delta_{+}}_{c}}$ with $\frac{T}{T_{c}}$ for different values of noncommutative parameter. In the left panel of the above Figure we have shown the variation of $\frac{\left<J\right>}{T^{\Delta_{+}}_{c}}$ with $\frac{T}{T_{c}}$ for $m^2 =0$ and $\Delta_{+}=1$. We have done the plots for different values of the noncommutative parameter. In the right panel, we have done the same plots for $m^2 =-3/16$ and $\Delta_{+}=3/4$. These plots are done for fixed values of the black hole mass $M=30/G_{4}$.}
	\label{fig 3}
\end{figure}
Substituting the value of $\chi$ from eq.\eqref{chi expand} in eq.\eqref{phi away critical}, we get
\begin{equation}
	\phi (z)=\lambda r_{+}(1-z)+\frac{\left<J\right>^2}{r^{2\Delta}_{+}}(\chi (0)+z\chi ^{'}(0)+\dots)~.
\end{equation}
Previously, we have seen that $\phi (z)$ is also represented by eq.\eqref{phi mu rho}, hence the coefficients of $z$ on both sides of the above equation can be compared, which leads to the following relation
\begin{equation}
	-\frac{\tilde{\rho}}{r_{+}}=-\lambda r_{+}+ \frac{\left<J\right>^2}{r^{2\Delta}_{+}}\chi ^{'}(0)~.
\end{equation}
In the above expression after substituting the value of $\chi ^{'}(0)$ from eq.\eqref{chi prime zero} one obtains
\begin{equation}
	\frac{\tilde{\rho}}{r^{2}_{+}}=\frac{\tilde{\rho}}{r^{2}_{+(c)}}\left[1+\frac{\left<J\right>^2}{r^{2\Delta}_{+}}\mathcal{A}_{\Delta}\right]~
\end{equation}
where we have used $\lambda=\frac{\tilde{\rho}}{r^{2}_{+(c)}}$~. Finally, using eq.(s)(\eqref{BH temp},\eqref{critical temp}) to replace $r_{+}$ and $r_{+(c)}$ in terms of the Hawking temperature ($T$) and the critical temperature ($T_c$), leads us to the following simplified expression
\begin{equation}
	\left<J\right>^{2}=\frac{(4\pi T_{c})^{2\Delta }}{\mathcal{A}_{\Delta}\left[3-\frac{4MG}{\Gamma(3/2)}\frac{e^{-\frac{(2MG)^{\frac{2}{3}}}{4\theta}}}{(4\theta)^{3/2}}\right]^{2\Delta }}\left(\frac{T_{c}}{T}\right)^{2}\left(1-\frac{T^2}{T^{2}_{c}}\right)~.
\end{equation}
We would like to mention that we are working in a regime which is away but close to the critical temperature, hence we can further approximate the term $\left(\frac{T_{c}}{T}\right)^{2}\left(1-\frac{T^2}{T^{2}_{c}}\right)$, which is as follows
\begin{equation}
	\left(\frac{T_{c}}{T}\right)^{2}\left(1-\frac{T^2}{T^{2}_{c}}\right)=\left(\frac{T_{c}}{T}\right)^{2}\left(1-\frac{T}{T_{c}}\right)\left(1+\frac{T}{T_{c}}\right) \approx 2\left(1-\frac{T}{T_{c}}\right)~.
\end{equation}
\begin{figure}[!h]
	\centering
	\begin{subfigure}[t]{0.5\textwidth}
		\centering
		\includegraphics[width=8.8cm]{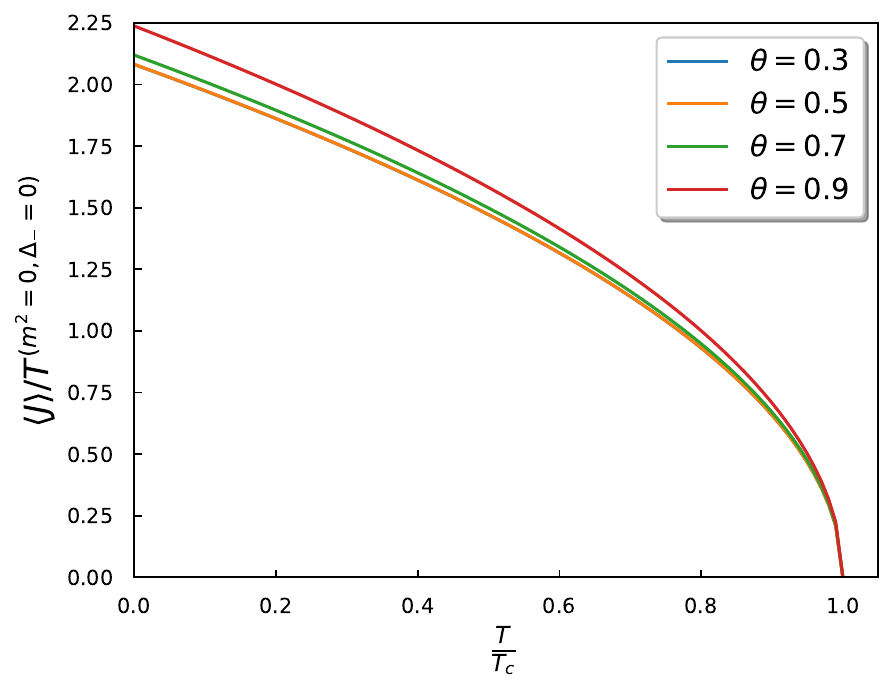}
		\caption{$m^2=0$ and $\Delta_{-}=0$}
		\label{con del minus 0}
	\end{subfigure}%
	~
	\begin{subfigure}[t]{0.5\textwidth}
		\centering
		\includegraphics[width=8.8cm]{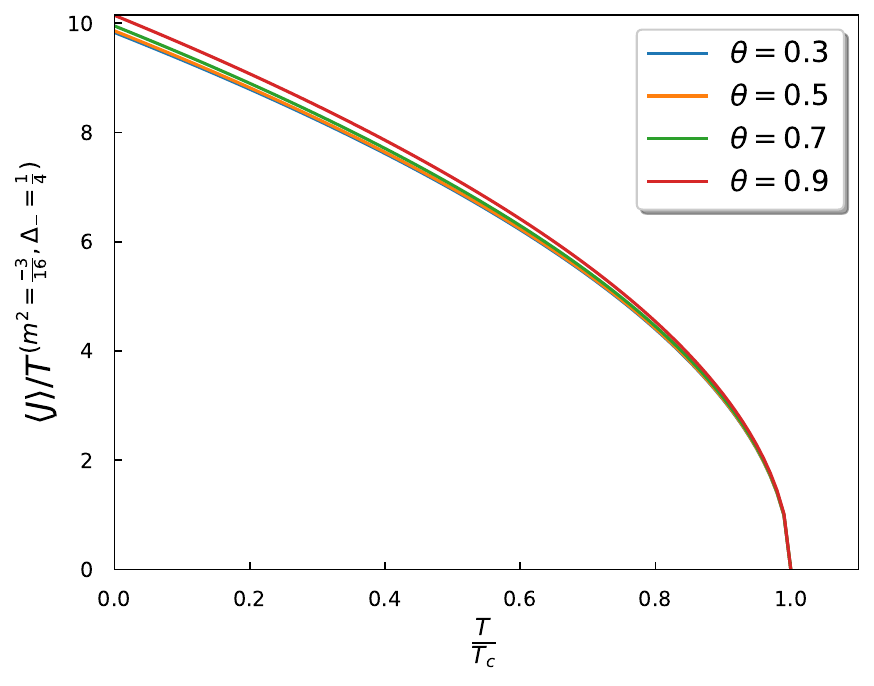}
		\caption{$m^2=-3/16$ and $\Delta_{-}=\frac{1}{4}$}
		\label{con del minus 1/4}
	\end{subfigure}
	\caption{Variation of the dimensionless condensate $\frac{\left<J\right>}{T^{\Delta_{-}}_{c}}$ with $\frac{T}{T_{c}}$ for different values of noncommutative parameter. In the left panel of the above Figure, we have shown the variation of $\frac{T_{c}}{\Tilde{\rho}^{1/2}}$ with black hole mass for $m^2 =0$ and $\Delta_{-}=0$. We have done the plots for different values of the noncommutative parameter. In the right panel, we have done the same plots for $m^2 =-3/16$ and $\Delta_{-}=1/4$. These plots are done for fixed values of the black hole mass $M=30/G_{4}$. }
	\label{fig 4}
\end{figure}
Hence, the final expression of the condensation operator becomes
\begin{equation}
		\left<J\right> = \beta T^{\Delta }_{c}\sqrt{1-\frac{T}{T_{c}}}
\end{equation}
where
\begin{equation*}
	\beta=\sqrt{\frac{2(4\pi)^{2\Delta }}{A_{\Delta}\left[3-\frac{4MG}{\Gamma(3/2)}\frac{e^{-\frac{(2MG)^{\frac{2}{3}}}{4\theta}}}{(4\theta)^{3/2}}\right]^{2\Delta }}}~.
\end{equation*}
We have also shown the above result graphically in Fig.(s)(\eqref{fig 3},\eqref{fig 4}). In Fig.(s)(\eqref{con del plus 1},\eqref{con del plus 3/4}), we have plotted the variation of the dimensionless condensate $\frac{\left<J\right>}{T^{\Delta_{+}}_{c}}$ with respect to $\frac{T}{T_{c}}$. We have done these plots for different values of the noncommutative parameter $\theta$ (that is, for $\theta = 0.3,~0.5,~0.7$ and $0.9$) and for a fixed value of the black hole mass $M=30/G_{4}$. Fig.\eqref{con del plus 1} is done for conformal dimension $\Delta_{+}=1$ and Fig.\eqref{con del plus 3/4} is done for conformal dimension $\Delta_{+}=\frac{3}{4}$. In both the plots, the curves in blue, orange, green and red correspond to the noncommutative parameter $\theta=0.3,~0.5,~0.7$ and $0.9$ respectively. \\
In Fig.(s)(\eqref{con del minus 0},\eqref{con del minus 1/4}), we have shown the variation of the dimensionless condensate $\frac{\left<J\right>}{T^{\Delta_{+}}_{c}}$ with respect to $\frac{T}{T_{c}}$. We done these plots for different values of the noncommutative parameter $\theta$ (that is, for $\theta = 0.3,~0.5,~0.7$ and $0.9$) and for a fixed value of the black hole mass $M=30/G_{4}$. Fig.\eqref{con del minus 0} is done for conformal dimension $\Delta_{-}=0$ and Fig.\eqref{con del minus 1/4} is done for conformal dimension $\Delta_{-}=\frac{1}{4}$. In both the plots, the curves in blue, orange, green and red corresponds to the noncommutative parameter $\theta=0.3,~0.5,~0.7$ and $0.9$ respectively.
\section{AC conductivity}\label{sec 6}
In this section, we will compute an expression for the AC conductivity of the noncommutative $p$-wave holographic superconductor using a self-consistent approach. To do so we will perturb the gauge field ($A_{\mu}$) along the boundary direction. Let us choose the following ansatz for the gauge field
\begin{equation}
	A_{\mu}=\left(0,0,\phi(r,t),0\right)
\end{equation}
where $\phi (r,t)=A(r)e^{-\iota\omega t}$~.\\
We would like to mention that although the gauge field is perturbed, we have not perturbed the massive vector field. Hence, we take the following ansatz for the vector field, given by
\begin{equation}
	\rho_{\mu}=\left(0,\rho(r),0,0\right)~.
\end{equation}
Following eq.\eqref{EOM F} and considering the above mentioned ansatz for matter and gauge field, the equation of motion for the radial part of the gauge field ($A(r)$) becomes
\begin{equation}
	A^{''}(r)+\frac{f^{'}(r)}{f(r)}A^{'}(r)+\left(\frac{\omega^{2}}{f^{2}(r)}-\frac{2\rho^{2}(r)}{r^{2}f(r)}\right)A(r)=0~.\label{gauge field perturbation eqn}
\end{equation}
To solve the above equation, we will move to the tortoise coordinate, given by
\begin{equation}
	r_{*}=\int \frac{dr}{f(r)} \implies dr_{*}=\frac{dr}{f(r)}~.\label{tortoise}
\end{equation}
Now writing eq.\eqref{gauge field perturbation eqn} in terms of the tortoise coordinate gives
\begin{equation}
	\frac{d^{2}A(r_{*})}{dr^{2}_{*}}+\left(\omega^{2}-V\right)A(r_{*})=0\label{gauge field in r star}
\end{equation}
where $V(r)=\frac{2\rho^{2}(r)f(r)}{r^2}$~. In terms of the $z$ coordinate, we can write
\begin{equation}\label{V}
	V(z)=2\rho^{2}(z)g(z)=2\rho^{2}(z)(1-z^{3}\Xi)~.
\end{equation}
For $V=0$, it is trivial to see that the solution of eq.\eqref{gauge field in r star} reads
\begin{equation}
	A\sim e^{-\iota\omega r_{*}}~.
\end{equation}
Before proceeding further, we must obtain an exact relation between the tortoise coordinate ($r_{*}$) and the inverse radial coordinate ($z$). This can be done by performing the integral in eq.\eqref{tortoise}. This reads
\begin{align}
	r_{*}=&\frac{1}{6\Xi^{1/3}r_{+}}\left(2\ln(1-\Xi^{1/3}z)-\ln(1+\Xi^{1/3}z+\Xi^{2/3}z^2)\right)\nonumber\\&-\frac{1}{\sqrt{3}r_{+}\Xi^{1/3}}\tan^{-1}\left(\frac{1+2z\Xi^{1/3}}{\sqrt{3}}\right)+c\label{r star with c}
\end{align}
where $c$ is the constant of integration. We have to evaluate the value of the constant $c$ such that our AdS boundary ($z=0$) is located at $r_{*}=0$ . Therefore, putting $z=0$ in the above equation gives us 
\begin{equation}
	c=\frac{\pi}{6\sqrt{3}\Xi^{1/3}r_{+}}~.
\end{equation}
After substituting this value of $c$ in eq.\eqref{r star with c} and simplifying, we finally obtain
\begin{equation}
	r_{*}=\frac{1}{6\Xi^{1/3}r_{+}}\left(2\ln(1-\Xi^{1/3}z)-\ln(1+\Xi^{1/3}z+\Xi^{2/3}z^2)\right)-\frac{1}{\sqrt{3}r_{+}\Xi^{1/3}}\tan^{-1}\left(\frac{\sqrt{3}z\Xi^{1/3}}{2+z\Xi^{1/3}}\right)~.
\end{equation}
Considering the leading order behaviour of $r_{*}$, we get
\begin{equation}
	r_{*}\sim \frac{1}{3\Xi^{1/3}r_{+}}\ln(1-\Xi^{1/3}z)=\ln(1-\Xi^{1/3}z)^{\frac{1}{3\Xi^{1/3}r_{+}}}~.
\end{equation} 
Hence, the leading order behaviour of the gauge field becomes
\begin{equation}
	A\sim (1-\Xi^{1/3}z)^{\frac{-\iota\omega}{3\Xi^{1/3}r_{+}}}~.
\end{equation} 
Now we will proceed to generalise the solution for the case $V\neq 0$. To do so, we will replace $V$ with $\left<V\right>$ in a self-consistent manner. In this case, we obtain 
\begin{equation}\label{A in V}
		A\sim (1-\Xi^{1/3}z)^{\frac{-\iota\sqrt{\omega^{2}-\left<V\right>^2}}{3\Xi^{1/3}r_{+}}}
\end{equation}
where $\left<V\right>$ is defined as
\begin{equation}
	\left<V\right>=\frac{\int dr_{*}V|A(r_{*})|^{2}}{\int dr_{*}|A(r_{*})|^{2}}~.
\end{equation}
Using the expressions of $V$ and $A$ from eq.(s)(\eqref{V},\eqref{A in V}) in the above equation, we obtain
\begin{align}
	\left<V\right>=&\frac{\int_{0}^{\infty}dz 2(1-z^{3}\Xi)F(z)^{2} \frac{\left<J\right>^{2}}{r^{\Delta}_{+}}e^{-2\sqrt{\left<V\right>}\sqrt{1-\frac{\omega^{2}}{\left<V\right>}}\frac{z}{r_{+}}}}{\int_{0}^{\infty}dz e^{-2\sqrt{\left<V\right>}\sqrt{1-\frac{\omega^{2}}{\left<V\right>}}\frac{z}{r_{+}}}}\nonumber\\
	&=\frac{\left<J\right>^{2}}{2^{2\Delta -1}}\frac{\Gamma(2\Delta +1)}{(\left<V\right>-\omega^{2})^{\Delta}}~
\end{align}
where $\Gamma(2\Delta+1)=\int_{0}^{\infty}t^{2\Delta}e^{-t}dt$.\\
We would like to mention that while obtaining the above result, we have assumed the fact that $F(z)\equiv 1$ and $r_{*}=\frac{-z}{r_{+}}$ close to the AdS boundary. In the low frequency regime, we obtain the following expression for $\left<V\right>$, which reads
\begin{equation}\label{V avg}
	\left<V\right>=\left(\frac{\left<J\right>^{2}}{2^{2\Delta-1}}\Gamma(2\Delta+1)\right)^{\frac{1}{\Delta+1}}~.
\end{equation}
Near the AdS boundary, we can expand the gauge field ($A(z)$) in eq.\eqref{A in V} as 
\begin{equation}
	A(z)=A(0)+z A^{'}(0)+\mathcal{O}(z^2)+\dots\label{A z}
\end{equation}
We also know that the gauge field near $z\to 0$, can also be expanded in the following manner
\begin{equation}
	A_{x}(z)=A^{(0)}_{x}+\frac{A^{(1)}_{x}}{r_{+}}z + \dots\label{A r}
\end{equation}
Comparing eq.(s)(\eqref{A z},\eqref{A r}), we get
\begin{equation}\label{A relations}
	A^{(0)}_{x}=A(0)~~, A^{(1)}_{x}=r_{+}A^{'}(0)~.
\end{equation}
From the gauge/gravity duality, the expression of conductivity is given by
\cite{PhysRevLett.101.031601,Hartnoll:2008kx,Horowitz:2008bn,Herzog:2009xv}
\begin{equation}
	\sigma(\omega)=\frac{\left<J\right>}{E_{x}}=-\frac{i A^{(1)}_{x}}{\omega A^{(0)}_{x}}~.
\end{equation}
Hence, using the relations from eq.\eqref{A relations}, the expression of conductivity is given by 
\begin{equation}\label{sigma def}
	\sigma(\omega)=-\frac{i r_{+}}{\omega}\frac{A^{'}(z=0)}{A(z=0)}~.
\end{equation}
Now substituting the expression of $A(z)$ from eq.\eqref{A in V} in the above equation, we get the following expression of conductivity
\begin{equation}
	\sigma(\omega)=\frac{1}{3\omega}\sqrt{\omega^{2}-\left<V\right>}~.
\end{equation}
Again in the low frequency limit, using the value of $\left<V\right>$ from eq.\eqref{V avg}, the expression of conductivity can be obtained as follows
\begin{equation}
	\sigma(\omega)=\frac{i}{3\omega}\left[\frac{\left<J\right>^{2}}{2^{2\Delta-1}}\Gamma(2\Delta+1)\right]^{\frac{1}{2(\Delta+1)}}~.
\end{equation}
It is clear from the above expression of AC conductivity that it has a pole of order one. Hence, the DC conductivity for the $p$-wave holographic superconductor model diverges in the $\omega\to 0$ limit. We will also like to mention that the effect of noncommutativity enters in our result through the condensation operator $\left<J\right>$ as it depends upon a factor $\beta$, which has different values for different values of black hole mass ($M$) and the noncommutative parameter ($\theta$). Although to observe the explicit dependence of conductivity on the noncommutative factor, we have to follow a more rigorous approach to solve the differential equation in eq.\eqref{gauge field perturbation eqn}. It should be mentioned that in the case of the self-consistent method, due to approximating the potential with its self-consistent average, we only get the leading order behaviour of the conductivity, and an explicit dependence of the noncommutative factor is missing.\\
In the upcoming subsections, we will calculate the conductivity of the holographic superconductor in a more rigorous manner. We will conduct our analysis primarily for two distinct scenarios. One in which $\rho_{-} = 0$, and another in which $\rho_{+} = 0$. It should also be mentioned that for simplicity we will restrict our calculations for $m^2 =0$, that is, for $\Delta_{+}=1$ and $\Delta_{-}=0$.  \\
\subsection{Case when $\texorpdfstring{\rho_{-}}{rho-}=0$}\label{6.2}
We will start our analysis by transforming eq.\eqref{gauge field perturbation eqn} in the $z$ coordinate, that reads
\begin{equation}\label{gauge pert in z}
	A^{''}(z)+\frac{g^{'}(z)}{g(z)}A^{'}(z)+\frac{1}{r^{2}_{+}}\left(\frac{\omega ^{2}}{g(z)^2}-\frac{\rho(z)^{2}}{g(z)}\right)A(z)=0~.
\end{equation}
Let us assume when $V\neq 0$, the gauge field takes the form
\begin{equation}
	A= (1-\Xi^{1/3}z)^{\frac{-\iota\omega}{3\Xi^{1/3}r_{+}}}G(z)~.
\end{equation}
Substituting this ansatz in eq.\eqref{gauge pert in z}, we obtain the following equation for $G(z)$
\begin{align}\label{eq G}
	&g(z)G^{''}(z)+\left(g^{'}(z)+\frac{2 i \omega g(z)}{3(-1+z \Xi^{1/3})r_{+}}\right)G^{'}(z)\nonumber\\&+\left(\frac{\omega^2}{g(z)r^{2}_{+}}+\frac{i\omega g(z)(i\omega +3 z\Xi ^{1/3}r_{+})}{9(-1+z\Xi^{1/3})^{2}r^{2}_{+}}+\frac{i\omega g^{'}(z)}{3 r_{+}-3z\Xi^{1/3}r_{+}}-\frac{\rho(z)^2}{r^{2}_{+}}\right)G(z)=0~.
\end{align}
Now, to study the properties of the function $G(z)$ in the low temperature limit, we will do a change of variable as $z=\frac{s}{b}$, where $b=\frac{\left<J\right>}{r_{+}}$. For low temperature, $r_{+}$ is small, hence $b$ is very large. Hence, taking the limit $b\to \infty$ implies working in the low temperature regime, 
where $G(z)$ is regular at the black hole horizon. Under the above-mentioned considerations, eq.\eqref{eq G} becomes 
\begin{equation}
	b^{2}G^{''}(s)+\frac{2i\omega b}{3 r_{+}}G^{'}(s)+\left(\frac{\omega^{2}}{r^{2}_{+}}+\frac{i\omega (i\omega+3\Xi^{1/3}r_{+})}{9 r^{2}_{+}}-\frac{\left<J\right>^{2}}{r^{4}_{+}}\frac{s^2}{b^2}\right)G(s)=0~.
\end{equation}
In the above equation, we have used the expression of $\rho(z)$ from eq.\eqref{ansatz rho} for $\Delta_{+}=1$.
At low frequency limit, that is, $\omega<<\left<J\right>$, the above equation can be approximated as follows
\begin{equation}
	G^{''}(s)-\frac{s^2}{\left<J\right>^2}G(s)=0~.
\end{equation} 
The solution of the above equation is given by\footnote{The actual solution comes in terms of parabolic cylinder D functions. In our analysis, we have taken terms up to $\mathcal{O}(z)$ of the series expansion of the parabolic cylinder D function. }
\begin{equation}\label{sol G}
	G(z)=c_{+}\left(\frac{\sqrt{\pi}}{2^{1/4}\Gamma(3/4)}-\frac{2^{3/4}\sqrt{\pi}z\sqrt{\left<J\right>}}{\Gamma(3/4)r_{+}}\right)+c_{-}\left(\frac{\sqrt{\pi}}{2^{1/4}\Gamma(3/4)}-\frac{i2^{3/4}\sqrt{\pi}z\sqrt{\left<J\right>}}{\Gamma(3/4)r_{+}}\right)~.
\end{equation}
To determine the constants $c_{+}$ and $c_{-}$ we must impose the condition that the function $G(z)$ regular at the black hole horizon ($z=1$), which leads to the following condition
\begin{equation}
	-3\Xi G^{'}(1)+\left(\frac{\omega^{2}(1+\Xi^{1/3})}{(1+\Xi^{1/3}+\Xi^{2/3})}-\frac{i \omega \Xi}{r_{+}(1-\Xi^{1/3})}-\frac{\left<J\right>^{2}}{r^{2}_{+}}\right)G(1)=0~.
\end{equation}
%\begin{equation}
	%6 %\sqrt{\frac{1}{\left<J\right>}}\Xi+\frac{\left<J\right>^{2}\left(\sqrt{\frac{1}{\left<J\right>}}\Gamma(3/4)-\frac{2(c_{+}+c_{-})\Gamm%a(5/4)}{c_{+}+i c_{-}}\right)}{\Gamma(3/4)r^{4}_{+}}=0
%\end{equation}
Substituting the expression of $G(z)$ from eq.\eqref{sol G} in the above equation, we get the following relation between the constants $c_{+}$ and $c_{-}$
\begin{equation}\label{c1,c2}
	\frac{c_{+}+c_{-}}{c_{+}+i c_{-}}=\frac{\frac{\Gamma(3/4)}{\left<J\right>^{5/2}}(\left<J\right>^{2}+3\Xi r^{4}_{+})}{2\Gamma(5/4)}~.
\end{equation}
Hence from the definition of the conductivity, we get
\begin{equation}
	\sigma(\omega)=\frac{-i r_{+}}{\omega}\frac{A^{'}(z=0)}{A(z=0)}=\frac{1}{3}+\frac{2 i \sqrt{\left<J\right>} (c_{+}+i c_{-})\Gamma(3/4)}{\omega(c_{+}+ c_{-})\Gamma(1/4)}~.
\end{equation}
Now substituting the value of $\frac{(c_{+}+i c_{-})}{(c_{+}+ c_{-})}$ from eq.\eqref{c1,c2} in the expression of the conductivity, we finally obtain 
\begin{equation}\label{AC conductivity rho plus 0}
	\sigma(\omega)=\frac{1}{3}+\frac{4i\left<J\right>^{3} \Gamma(5/4)}{\omega\Gamma(1/4)(\left<J\right>^{2}+3\Xi r^{4}_{+})}~.
\end{equation}
From the above expression of conductivity, it is clear that the AC conductivity has a pole of order one in the frequency domain. Also, it can be seen from our analysis that the noncommutative corrections enter our results as a correction to the leading order of the AC conductivity. \\
One can also check the $\theta\to 0$ limit for the expression of the AC conductivity in eq.\eqref{AC conductivity rho plus 0}. From the expression of $\Xi$ in section \eqref{sec 4}, it is clear that at $\theta\to 0$, $\Xi\approx 1$, using this value of $\Xi$ in eq.\eqref{AC conductivity rho plus 0}, we get the following expression of AC conductivity
\begin{equation}
    \sigma(\omega)\mid_{\theta\to 0}\approx\frac{1}{3}+\frac{4i\left<J\right>^{3} \Gamma(5/4)}{\omega\Gamma(1/4)(\left<J\right>^{2}+3 r^{4}_{+})}~.
\end{equation}
One can clearly see that just like the noncommutative case, in the $\theta\to 0$ limit, the AC conductivity also has a first-order pole at $\omega =0$.
\\
\noindent We already know that if the band gap energy is $E_{g}$, then it is related to the real part of the conductivity in the following manner
\begin{equation}\label{Re sigma 0}
	Re[\sigma(\omega=0)]=e^{-\frac{E_{g}}{T}}~.
\end{equation}
Hence the band gap energy at temperature $T$ is given by
\begin{equation}
	E_{g}= \frac{r_{+}\ln(3)}{4\pi}\left[3-\frac{4MG}{\Gamma(3/2)}\frac{e^{-r^{2}_{+}/4\theta}}{(4\theta)^{3/2}}\right]~.
\end{equation}
The above expression represents the band gap energy of our noncommutative holographic superconductor model when $\rho_{-} = 0$.
It is also clear from the above expression that in this noncommutative model (for $m^2 =0$ and $\Delta_{+}=1$), the band gap energy of the holographic superconductor depends upon both the black hole mass ($M$) and the non-commutative parameter ($\theta$).
\subsection{Case when $\texorpdfstring{\rho_{+}}{rho+}=0$}
In this subsection, we will calculate the DC conductivity of the holographic superconductor considering $\rho_{+}=0$ and $\Delta_{-}=0$. We will start from eq.\eqref{eq G} and substitute the value of $\rho (z)$ from eq.\eqref{ansatz rho} for $\Delta =\Delta_{-} =0$. This leads to 
\begin{align}\label{G rho plus 0}
	&g(z)G^{''}(z)+\left(g^{'}(z)+\frac{2 i \omega g(z)}{3(-1+z \Xi^{1/3})r_{+}}\right)G^{'}(z)\nonumber\\&+\left(\frac{\omega^2}{g(z)r^{2}_{+}}+\frac{i\omega g(z)(i\omega +3 z\Xi ^{1/3}r_{+})}{9(-1+z\Xi^{1/3})^{2}r^{2}_{+}}+\frac{i\omega g^{'}(z)}{3 r_{+}-3z\Xi^{1/3}r_{+}}-\frac{\left<J\right>^2}{r^{2}_{+}}\right)G(z)=0~.
\end{align}
As mentioned before in subsection \eqref{6.2}, we will do a change of variable, that is, $z=\frac{s}{b}$, where $b=\frac{\left<J\right>}{r_{+}}$. Now to study the properties $G(z)$ in the low temperature regime, we will also take the limit $b \to \infty$ in eq.\eqref{G rho plus 0}. This gives 
\begin{equation}
	b^{2}G^{\prime\prime}(s)-\frac{2 i \omega }{3 r_{+}}G^{\prime}(s)+\left(\frac{8 \omega^2}{r^{2}_{+}}-\frac{\left<J\right>^2}{r^{2}_{+}}\right)G(s)=0~.
\end{equation}
For low frequency solutions of $G(z)$, we will take $\omega << \left<J\right>$. Under this consideration, the above equation becomes
\begin{equation}
	G^{\prime\prime}(s)-G(s)=0~.
\end{equation} 
The solution of the above equation is given by
\begin{equation}
	G(z)=c_{1}e^{-\frac{\left<J\right>}{r_{+}}z}+c_{2}e^{\frac{\left<J\right>}{r_{+}}z}~.
\end{equation}
In order to determine the constants $c_{1}$ and $c_{2}$, we must impose the condition that the function $G(z)$ regular at the black hole horizon ($z=1$), which leads to the following condition
\begin{equation}
	-3\Xi G^{'}(1)+\left(\frac{\omega^{2}(1+\Xi^{1/3})}{(1+\Xi^{1/3}+\Xi^{2/3})}-\frac{i \omega \Xi}{r_{+}(1-\Xi^{1/3})}-\frac{\left<J\right>^{2}}{r^{2}_{+}}\right)G(1)=0~.
\end{equation}
If we substitute the expression of $G(z)$ from eq.\eqref{sol G} in the above equation, we get the following relation between the constants $c_{1}$ and $c_{2}$, that reads
\begin{equation}
	\frac{c_{1}}{c_{2}}=e^{-2 b}\frac{-b^2 r^{2}_{+}(\Xi-1)+r_{+}(3b r_{+} (\Xi -1 )\Xi +2 i (1+\Xi^{1/3}+\Xi^{2/3})\Xi \omega -\omega^{2}r_{+})}{b^2 r^{2}_{+}(\Xi-1)+r_{+}(3b r_{+} (\Xi -1 )\Xi -2 i (1+\Xi^{1/3}+\Xi^{2/3})\Xi \omega +\omega^{2}r_{+})}~.
\end{equation}
In the low temperature limit, the above expression can be approximated as 
\begin{equation}\label{c1 / c2 }
	\frac{c_1}{c_2}\approx-e^{-2b}\left(1+\frac{4\omega^{2}}{b^{4}r^{2}_{+}}\Theta(\Xi)^2\right)
\end{equation}
where $\Theta(\Xi)=\Xi (1+\Xi^{1/3}+\Xi^{2/3})$.\\
Again, from the definition of the conductivity in eq.\eqref{sigma def}, we get
\begin{equation}
	\sigma (\omega)=\frac{-i}{\omega} \left[\left<J\right>\frac{\left(\frac{c_1}{c_2}-1\right)}{\left(\frac{c_1}{c_2}+1\right)}+\frac{i \omega}{3}\right]~.
\end{equation}
If we substitute the value of $\frac{c_1}{c_2}$ from eq.\eqref{c1 / c2 }, the above equation of conductivity for low temperature and low frequency is approximately given by
\begin{equation}\label{AC conductivity rho minus 0}
	\sigma (\omega) \approx \frac{- i \left<J\right>}{\omega}\left[1+2e^{-2b}\left(1+\frac{4\omega^{2}}{b^{4}r^{2}_{+}}\Theta(\Xi)^2\right)\right]+\frac{1}{3}~.
\end{equation}
From the above expression of conductivity, it is clear that the result has contributions from the noncommutative parameter, which enters in our result through $\Theta (\Xi)$. We would also like to highlight that the impact of noncommutativity on the expression for DC conductivity is minimal. This is due to the fact that the term associated with the noncommutative parameter is accompanied by a prefactor of $\frac{1}{b^4}$, which significantly reduces its contribution. Similarly to the previous case, the AC conductivity exhibits a first-order pole.\\
For this case also, we can take the commutative limit (that is $\theta\to 0$) for the expression of the AC conductivity given in eq.\eqref{AC conductivity rho minus 0}. As we have mentioned in the previous subsection that in the $\theta\to 0$ limit, $\Xi \approx 1$, thereby $\Theta(\Xi)\approx 3$. Using these values of $\Xi$ and $\Theta(\Xi)$ in eq.\eqref{AC conductivity rho minus 0}, we get
\begin{equation}
    \sigma (\omega)\mid_{\theta\to 0} \approx \frac{- i \left<J\right>}{\omega}\left[1+2e^{-2b}\left(1+\frac{36\omega^{2}}{b^{4}r^{2}_{+}}\right)\right]+\frac{1}{3}~.
\end{equation}
In the commutative case also we can see that the AC conductivity has a first order pole in the frequency domain.
\\
 Now we will proceed to calculate the band gap energy of this holographic superconductor model. To do so, we will first calculate the real part of the AC conductivity, which reads
 \begin{equation}
 	Re[\sigma (\omega)]=\frac{1}{3}~.
 \end{equation}
 %\frac{-8 i \left<J\right> \omega \Theta(\Xi)^{2}}{b^{4}r^{2}_{+}}e^{-2b}+
Hence, $Re[\sigma (\omega = 0)]=\frac{1}{3}$. Now using eq.\eqref{Re sigma 0}, we get the band gap energy to be
\begin{equation}\label{bandgap final}
	E_{g}= \frac{r_{+}\ln(3)}{4\pi}\left[3-\frac{4MG}{\Gamma(3/2)}\frac{e^{-r^{2}_{+}/4\theta}}{(4\theta)^{3/2}}\right]~.
\end{equation}
It is interesting to observe that, although the expressions for DC conductivity differ in the cases of $\rho_{+} = 0$ and $\rho_{-} = 0$, their real parts are identical, which leads to the same expressions for the band gap energy. Some comments on the physical effects of the noncommutative parameter on the band gap energy are in order. As mentioned earlier, the noncommutative parameter $\theta$ implements a physical length scale below which fields and particles can not be localised \cite{Nicolini:2005vd,Nicolini:2008aj}. In AdS/CFT, the region close to the black hole horizon is referred to as the IR limit of the theory. The noncommutative parameter smears the mass distribution of the black hole and matter fields close to the horizon; therefore, $\theta$ modifies the IR behaviour of the theory and effectively introduces an IR cutoff. Also the presence of noncommutativity ensures that the charge of the black hole is no longer sharply localised at the horizon but smears through the geometry. This smearing of charge affects the bulk gauge field profile $\phi(r)$. Therefore it also affects the band gap energy. From eq.\eqref{bandgap final} it is clear that the increasing value of $\theta$ suppress the value of band gap energy.\\
In the limit $\theta\to0$, the second term in the above expression of the band gap energy is exponentially suppressed. This leads to the following expression for the band gap energy
\begin{equation}
    E^{(\theta\to 0)}_{g}\approx \frac{3r_{+}\ln(3)}{4\pi}~.
\end{equation}
Now the temperature of the black hole in the commutative limit ($\theta\to 0$) is given by $T^{(\theta\to0)}=\frac{3 r_{+}}{4\pi}$.
Therefore, we get 
\begin{equation}
    \frac{E^{(\theta\to 0)}_{g}}{T^{(\theta\to0)}}=\ln{3}\approx 1.098~.
\end{equation}
We would like to mention that although there are previous studies on the commutative case of Maxwell $p$-wave holographic superconductors, none of them  have computed an analytical expression for the above ratio\footnote{For BCS theory, which is a theory of weakly coupled superconductors, the  ratio between the band gap energy and the critical temperature is $1.736$\cite{PhysRevLett.101.031601}.}. 
\section{Conclusion}\label{sec 7}
In this paper, we have studied the Maxwell $p$-wave holographic superconductor model in a noncommutative AdS$_4$ background for Maxwell electrodynamics. For simplicity, we have performed all the calculations in the probe limit, which ensures no back reaction on the spacetime geometry due to the matter and gauge field. At first, we have discussed about the Maxwell $p$-wave holographic superconductor model in presence of noncommutativity. Then we have discussed about the asymptotic and near critical temperature behaviour of the matter and gauge fields, using these we have obtained an analytical expression for the critical temperature analytically by applying the St\"{u}rm-Lioville eigenvalue approach. We have calculated the values of the critical temperature for four different cases (that is, for $\Delta_{+}=1,\frac{3}{4}$ and $\Delta_{-}=0,\frac{1}{4}$). With the expression of critical temperature in hand, we have then graphically shown the variation of critical temperature with respect to black hole mass for different values of the noncommutative parameters. From the graphs, it is clear that increasing the value of the noncommutative parameter ($\theta$) actually makes the condensation harder to form. A possible reason for this is that the noncommutative parameter introduces an IR cutoff and smears both the black hole mass and charge. The features for this model are very much similar in nature to the noncommutative $s$-wave models studied in \cite{Pramanik:2014mya,Ghorai:2016qwc}. We would also like to mention that the previous study on the noncommutative $s$-wave model \cite{Pramanik:2014mya} deals with a scalar field with mass $m^2 =-2$ and conformal dimension $\Delta_+=1$. Also this study uses different values of black hole masses to obtain the values of the critical temperature, which is completely different compared to the values used in our analysis. The extension of this model in the Born-Infeld electrodynamics \cite{Ghorai:2016qwc} uses conformal dimension ($\Delta_+=3$) and spacetime dimension five. The non-Abelian $p$-wave model \cite{Gangopadhyay:2012am} also deals with different conformal dimensions. Therefore, the physical quantities in our paper have different numerical values compared to those in the previous works. Although we have provided a side by side comparison of our critical temperature values in our model for $m^2 =0$, $\Delta_+ =1$ and bulk spacetime dimension $d=4$ with the noncommutative $s$-wave model results \cite{Ghorai:2016qwc} for $m^2$=-3, $\Delta_+ =3$ and spacetime dimensions $d=5$. From this comparison, we have seen that for both the models, increasing the value of the noncommutative parameter suppresses the value of the critical temperature, thus making the condensate harder to form. We have also provided a small comparison between our results in Table \eqref{tab 1} for $\theta=0.5$ and $M=30/G_4$ with the value of $\frac{T_{c}}{\sqrt{\Tilde{\rho}}}$ for $\theta=0.5$ and $M\approx\frac{28.2}{G_4}$ in \cite{Pramanik:2014mya}. This comparison also shows the similarity between the features of $s$ and $p$-wave models. We would also like to mention that for higher black hole masses, all the graphs corresponding to various noncommutative parameters tends to a fixed value of critical temperature of the holographic superconductor. This suggests that for larger black hole masses, the effect of noncommutativity do not play a significant role in the properties of the superconductor. The above observations are true for all the cases corresponding to $\Delta_{+}=1,\frac{3}{4}$ and $\Delta_{-}=0,\frac{1}{4}$.\\
To derive the expression of the condensation operator, we have studied the properties of gauge and matter fields slightly away from the critical temperature.We have derived a general expression for the condensation operator through analytical calculations and identified the critical exponent as $\frac{1}{2}$. Additionally, we have demonstrated how the condensation operator varies with temperature for various selected values of the conformal dimensions discussed earlier.\\
Finally, we have calculated the AC conductivity of the holographic superconductor. To do so we have perturbed the gauge field along the boundary direction. Then the differential equation for the radial part of the gauge field is solved to obtain the expression of the AC conductivity. We initially used the self-consistent method, substituting the potential with its self-consistent average, which allowed us to derive an expression for conductivity applicable to any arbitrary conformal dimension in the low frequency and low temperature limit. We have also noticed that the expression of conductivity has a pole of order one in the frequency domain. We would also like to highlight that the dependence of the noncommutative factor is incorporated into the conductivity expression via the condensation operator. The explicit dependence of noncommutative factor is missing in the case of the self-consistent method because we have to approximate the potential with its self-consistent average. This approach only provides the leading-order behaviour of the conductivity. To investigate the direct relationship between AC conductivity and the noncommutative factor, we have adopted a more rigorous approach for solving the gauge field perturbation equation. We have done the calculations in the low temperature and low frequency limit for $m^{2}=0$ (that is for $\Delta_{+}=1$ and $\Delta_{-}=0$). In both cases, we observe an explicit dependence of the noncommutative factor on the conductivity expression. We have also calculated the values of AC conductivity in the commutative limit for both the cases with $\rho_+ =0$ and $\rho_-=0$. For the commutative case also, we have seen the presence of a first order pole in the frequency domain. It is important to highlight that, although the mathematical forms of the conductivity expressions are different for both instants, their real parts are identical in the zero frequency limit, resulting in the same expression for the band gap energy. The dependence of the noncommutative factor on band gap energy is also prominent in our results. As a future direction of this work one can see the effects of nonlinear electrodynamics in this $p$-wave holographic superconductor model along with noncommutativity and observe the effects of the noncommutative factor and nonlinear parameter in the condensation and AC conductivity. \\

\section*{Appendix}
\section*{Calculation of \texorpdfstring{$\lambda_{\alpha_{min}}$}{lambda-alpha-min}}
In order to calculate the numerical value of $\lambda_{\alpha_{min}}$, one needs to go through the following steps.
\begin{enumerate}
    \item Using the ansatz for $F_{\alpha}(z)$ in eq.\eqref{lamda square general}, we will take the first order derivative of $\lambda^2$ with respect to $\alpha$.
    \item Now the expression for $\frac{d \lambda^2}{d \alpha}$ will be set to zero to find the roots of $\alpha$. These roots are calculated using numerical root finding.
    \item Among several roots of alpha, we will now need to find the root that minimises the value of $\lambda^2$. This can be done by checking the sign of $\frac{d^2\lambda^2}{d\alpha^2}$. The root of $\alpha$ for which $\frac{d^2\lambda^2}{d\alpha^2}>0$ gives the minimum value corresponding to $\lambda^2$.
    \item With the required root of $\alpha$ in hand, we can calculate the corresponding minimum value of $\lambda^2$ using eq.\eqref{lamda square general}. 
\end{enumerate}
Now we will perform the calculation of $\lambda_{\alpha_{min}}$ for $\theta=0.3$ and $M=10/G_4$ for case-I in section-\ref{sec 4}. For these values of $\theta$ and $M$, we can use the form of $p(z)$, $q(z)$ and $r(z)$ from eq.\eqref{pqr case1} and evaluate the integrals in eq.(s)(\eqref{I1},\eqref{I2}).
For $\theta=0.3$ and $M=10/G_4$ the expression of $I_1$ and $I_2$ are given as follows
\begin{align}
    I_1 &=\int_{0}^{1}dz\Big(4z^4 (1-1.00652z^3)\alpha^2 +3.01955 z^3 (1-\alpha z^2)^2\Big)\nonumber\\&=0.754887 - 1.00652 \alpha + 0.674186 \alpha^2\\
    I_2 &=\int_{0}^{1}dz\frac{(1-z)^2 z^2 (1-\alpha z^2)^2}{1-1.00652z^3}=0.0439622-0.0307884\alpha +0.00768053\alpha^2~.
\end{align}
Using these expressions of $I_1$ and $I_2$ from the above equation, the expression of $\lambda^2$ in terms of $\alpha$ follows from eq.\eqref{lamda square general}, which reads
\begin{equation}\label{lambda2 alpha appendix}
    \lambda^2=\frac{0.754887 - 1.00652 \alpha + 0.674186 \alpha^2}{0.0439622-0.0307884\alpha +0.00768053\alpha^2}~.
\end{equation}
Now in order to minimize $\lambda^2$, we will differentiate eq.\eqref{lambda2 alpha appendix} with respect to $\alpha$, which gives
\begin{align}
    \frac{d\lambda^2}{d\alpha}&=\frac{-1.00652 + 1.34837\, \alpha}{
 0.0439622 - 0.0307884\, \alpha + 
 0.00768053\, \alpha^2}
\nonumber\\&-
\frac{
  (-0.0307884 + 0.0153611\, \alpha)\, (0.754887 - 1.00652\, \alpha + 0.674186\, \alpha^2)
}{
  \left(0.0439622 - 0.0307884\, \alpha + 0.00768053\, \alpha^2\right)^2
}~.
\end{align}
The roots of the above equation are obtained using the numerical root-finding method in MATHEMATICA $13.3$. The three roots of the above equation are $6.18653\cross 10^{16}$, $3.14809$ and $0.512256$. The errors in the numerical roots are negligibly small. Now we need to find out among these three roots which one gives the minimum value of $\lambda^2$. To do so, we need to check the sign of $\frac{d^2\lambda^2}{d\alpha^2}$. For these roots, the value of $\frac{d^2\lambda^2}{d\alpha^2}$ is given by
\begin{align}
    &\frac{d^2\lambda^2}{d\alpha^2}\mid_{\alpha=6.18653\cross 10^{16}}= -4.37906\cross 10^{-47}<0\nonumber\\
    &\frac{d^2\lambda^2}{d\alpha^2}\mid_{\alpha=3.14809}=-64.0401<0\nonumber\\
    &\frac{d^2\lambda^2}{d\alpha^2}\mid_{\alpha=0.512256}=37.6319>0~.
\end{align}
Therefore, $\alpha=0.512256$ is the root corresponding to the minimum value of $\lambda^2$, Hence using this value of $\alpha$ in eq.\eqref{lambda2 alpha appendix}, we get $\lambda_{\alpha_{min}}=13.7788$. This value of $\lambda_{\alpha_{min}}$, has been used to find the value of $\frac{T_c}{\Tilde{\rho^{1/2}}}$ using eq.\eqref{critical temp}.
\section*{Acknowledgment}
SP would like to thank SNBNCBS for Junior Research Fellowship. The authors also thank the anonymous referees for their useful comments.
\bibliographystyle{hephys}
\bibliography{citation-hsc.bib}
\end{document}